\documentclass[12pt]{article}
\usepackage{graphicx}
\usepackage{color}
\newcommand{\beq}{\begin{equation}}
\newcommand{\eeq}{\end{equation}}
\newcommand{\bea}{\begin{eqnarray}}
\newcommand{\eea}{\end{eqnarray}}

\newcommand{\gsim}{\lower.7ex\hbox{$
\;\stackrel{\textstyle>}{\sim}\;$}}
\newcommand{\lsim}{\lower.7ex\hbox{$
\;\stackrel{\textstyle<}{\sim}\;$}}

\newcommand{\eod}{\end{document}}

\newcommand{\epem}{$e^+e^-$}

\definecolor{verm}{rgb}{0.8,0.1,0.0}

\include{chref}

\begin{document}
\thispagestyle{empty}
\vspace*{-22mm}

\begin{flushright}

INFN-20-01/LNF\\
UND-HEP-19-BIG\hspace*{.08em}03\\
Version 2.5 
\end{flushright}

\vspace*{0.2mm}

\begin{center}

{\Large {\bf 2019/20 Lessons from $\tau (\Omega_c^0)$ \& $\tau (\Xi_c^0)$  and {\bf CP} asymmetry in charm decays}}

\vspace*{3.mm}

{\bf S. Bianco $^a$, I.I.~Bigi $^b$}\\
\vspace{1mm}
$^a$ {\em INFN, Laboratori Nazionali di Frascati, Frascati (Rome), I-00044, Italy}

$^b$ 
{\em Department of Physics, University of Notre Dame, Notre Dame, IN 46556, USA} 

\vspace*{-.8mm}

{\sl email addresses: stefano.bianco@lnf.infn.it, ibigi@nd.edu}

\vspace*{3mm}

{\bf Abstract} 

\end{center}

\vspace*{-3.0mm}

\noindent 
Our 2003  "Cicerone" had discussed charm dynamics  with different directions and levels \cite{CICERONE}. 
Here we focus on two items, where the `landscape' has changed sizably.  
(A) The lifetimes \& semi-leptonic decays of charm hadrons show the impact of non-perturbative QCD and to which degree one can apply 
Heavy Quark Expansion (HQE) for charm hadrons. It is more complex as we have learnt from 2019/20 data. 
(B) {\bf CP} asymmetry has been established in 2019 \cite{LHCbGuy}:  
$\Delta A_{\bf CP} \equiv  A_{\bf CP}(D^0 \to K^+K^-) - A_{\bf CP}(D^0 \to \pi^+\pi^-) $=$\, -\, (1.54 \pm 0.29 ) \cdot 10^{-3}$ is 
quite an achievement by the LHCb collaboration! 
Our community is at the beginning of a long travel for fundamental dynamics. Can the SM account for these? 
We discuss the assumptions that were made up to 2018 data and new conclusions from 2019/20 ones. 
We need more data; however, one has to discuss correlations between different transitions.  
We give an {\bf Appendix} what we have learnt for large {\bf CP} asymmetry in $K_L \to \pi^+\pi^-e^+e^-$.

\vspace{5mm}

\tableofcontents

\section{Introduction} 
\label{INTRO}

The goal of "Cicerone"\cite{CICERONE} was and still is to reach a large audience including graduate students: first it `paints' the `landscape' 
of charm hadrons and then goes deeper inside the fundamental dynamics and refined tools in general. 
However, we do not discuss spectroscopy in this paper, although the topic is important and relevant progress has been made by our community. 
We focus on two items and discuss that with colleagues who work on them (or close to it) both on the 
theoretical or experimental  side. 
\begin{itemize}
\item
How can one improve the predictions for the lifetimes and semi-leptonic branching ratios of charm hadrons from HQE (Heavy Quark Expansion)  
\footnote{Often it is said the words of `HQE' and `HQET' mean the same; however, we disagree.}? 
\label{LIFEHQE}
\item
{\bf CP} asymmetry in charm hadrons has been established in 2019 in one case: $\bf CP$ asymmetry in 
$D^0 \to K^+K^-$ vs. $D^0 \to \pi^+\pi^-$ \cite{LHCbGuy}. 
\label{CHARMCPV}
\end{itemize} 
Our community is in the beginning of a long `travel' through fundamental dynamics; in particular, the SM 
predicts small, but not zero values for {\bf CP} violation in singly Cabibbo suppressed (SCS) transitions of charm hadrons. One has to find that in other ones, 
namely in $D^+$ \& $D^+_s$ and charm baryons decays at least in $\Lambda_c^+$ and hopeful in $\Xi_c^+$ \& $\Xi_c^0$  ones. 
For practical reasons one first focuses on 2-body final states (FS). However, our community has to find {\bf CP} asymmetries in 3- \& 4-body FS. 
Most of non-leptonic decays of charm hadrons  are given 
by 3 \& 4 hadrons like pions, kaons, $\eta$ \& $\eta^{\prime}$ and $p$, $\Lambda$ etc.; they are {\em not} backgrounds. 
It is crucial to understand the underlying dynamics: the SM produces these data, or it is SM plus New Dynamics (ND). 
These two items are not the same, but they are connected due to QCD, as we will explain below. 

We give a very short introduction to help a reader to understand our points.  
The beauty quark is heavy compared to 1 GeV; thus one can apply HQE to deal with non-perturbative QCD; 
it has a good record for lifetimes, branching ratios, oscillations and even {\bf CP} asymmetries.   
The weak widths of $H_Q$ with single heavy quark $Q$ are described by Operator Product Expansion (OPE):  
\beq
\Gamma (H_Q \to f) = \frac{G_F^2 m_Q^5(\mu)}{192\pi^3}|V_{\rm CKM} |^2 \left[ A_0 +A_2/m^2_Q + A_3/m^3_Q +{\cal O}(1/m_Q^4) \right]_{(\mu)}  \; . 
\label{OPEGEN}
\eeq
$A_n$ are numbers containing phase space factors, short-distance coefficients appearing in OPE and hadronic expectation values of local operators 
${\cal O}_n$ of dimension $n +3$, namely $\langle H_Q |{\cal O}_n |H_Q \rangle$. 
Quantum field theory (QFT) had told us: (1) $A_0 \neq 0$ is the same for $Q$ quark except phase space factors and the value of $\langle H_Q |\bar QQ |H_Q \rangle$. 
(2) In HQE $A_1$ is zero. (3) The non-zero values of $A_2$ are different for heavy baryons, while basically they are still zero for heavy mesons. 
(4) $A_3$ give different non-zero values for heavy mesons, heavy baryons and also for the connections of heavy baryons \& heavy mesons. 
(5) Our community has worked beyond $A_3$, namely $A_4$; however, we will not discuss that here. 

As usual, one  introduces an auxiliary energy scale $\mu$ with $\Lambda _{\rm QCD}  \ll   \mu \ll m_Q$ for the operators and for the matrix elements to combine 
them to get observables that do not depend on $\mu$. 
It is indicated in this Eq.(\ref{OPEGEN}); for practical reasons one gets $\mu \sim$ 1 GeV.

To discuss the transitions of charm hadrons we summarize the status of SM predictions vs. 2018 data and then discuss the sizable differences in 2019/20 data 
\footnote{We talk about HQE with only two examples, namely beauty \& charm hadrons. On the other hand, one can look at hyperfine splitting with four `actors': 
$M^2 (\rho) - M^2 (\pi) \sim 0.59$ (GeV)$^2$, $M^2 (K^*) - M^2 (K) \sim 0.56$ (GeV)$^2$, 
$M^2 (D^*) - M^2 (D) \sim 0.53$ (GeV)$^2$ \& $M^2 (B^*) - M^2 (B) \sim 0.48$ (GeV)$^2$. It is `luck' -- or we missed something?}.  
There are two obvious differences between the 2018 and 2019/20 data. 
The history plots of the measurements of the lifetime of $\Omega_c^0$: following  the PDG style, they are shown in  {\bf Fig.\ref{FIG:HISTORYLIFE}}:  
\begin{figure}[h!]
\begin{center}
\includegraphics[width=8cm]{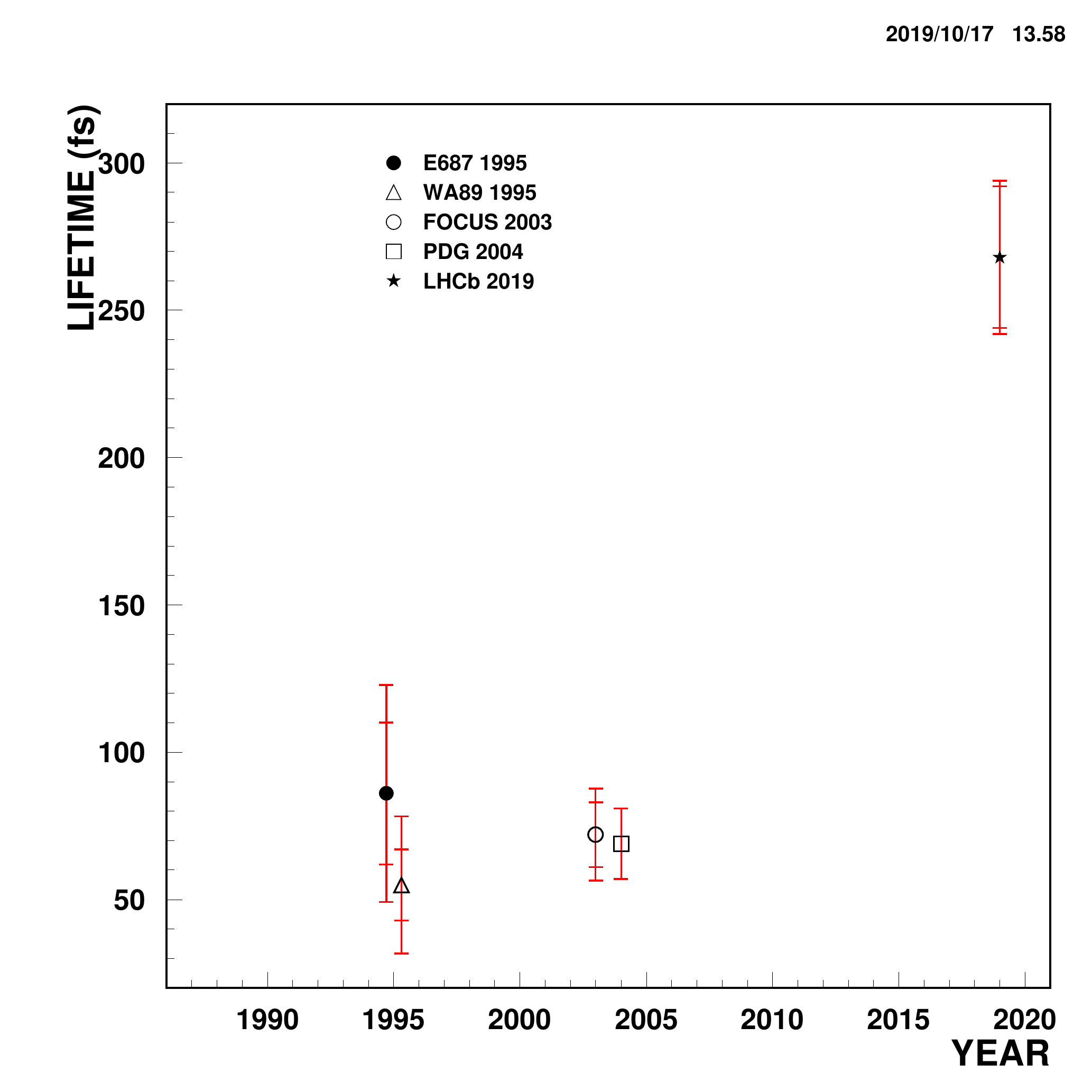}
\end{center}
\vspace{-0.5cm}
\caption{History plot for $\Omega_c$ lifetimes.}
\label{FIG:HISTORYLIFE}
\end{figure}
It had been seemed the data are controlled; however the lifetime of $\Omega_c^0$ measured by the LHCb collaboration is very different! 
As we will discuss below in {\bf Sect.\ref{BARYONSLIFE}}, it has changed our understanding the underlying dynamics of non-perturbative QCD. 

\begin{figure}[h!]
\begin{center}
\includegraphics[width=8cm]{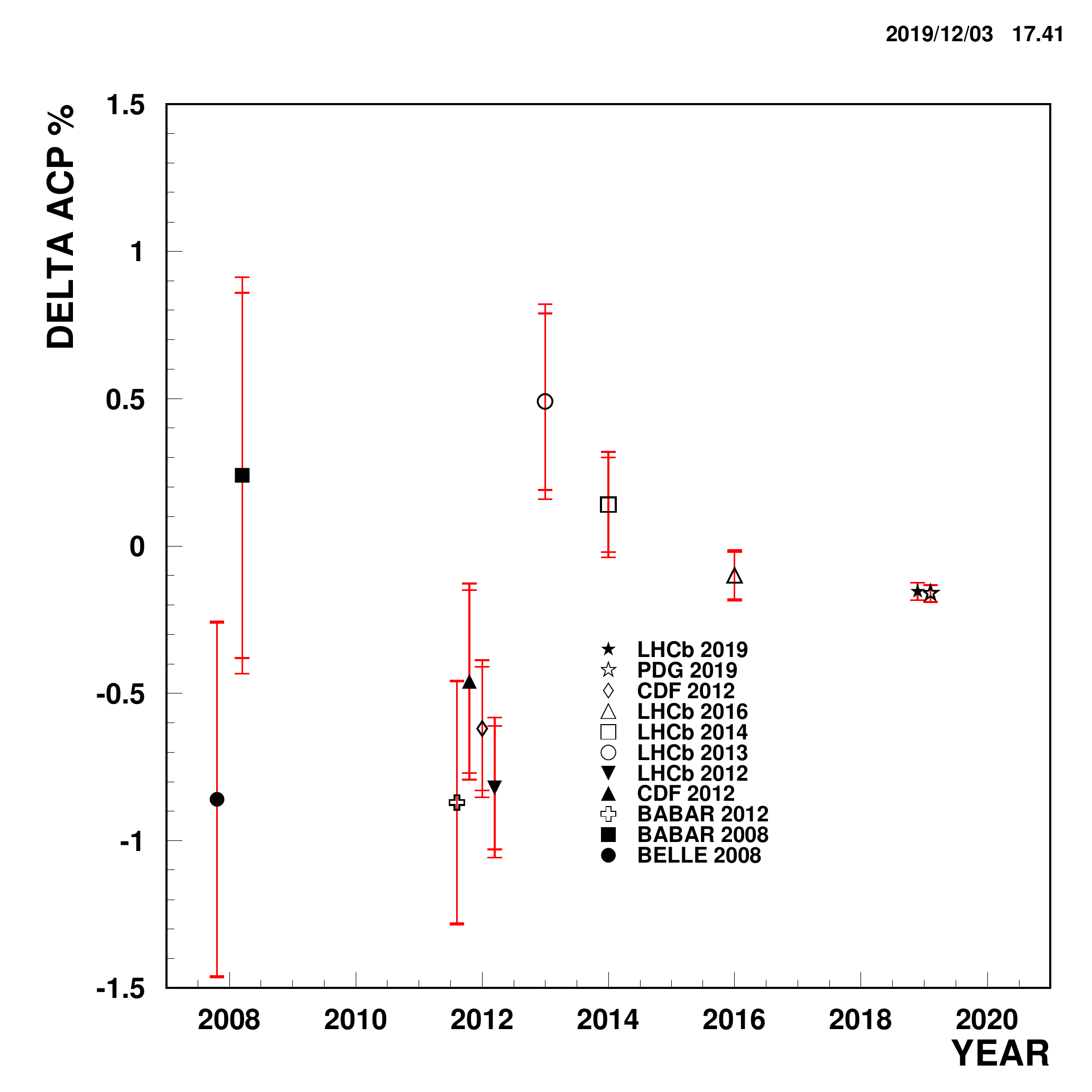}
\end{center}
\vspace{-0.5cm}
\caption{History plot for CP asymmetry differences $D^0 \to K^+K^-$/$\pi^+\pi^-$.}
\label{FIG:HISTORYCPASYM}
\end{figure}

The situation is very different for {\bf CP} asymmetries in $D^0 \to K^+K^-/\pi^+\pi^-$, see {\bf Fig.\ref{FIG:HISTORYCPASYM}}. 
The result of the LHCb analyses is indeed exciting \cite{LHCbGuy}; thus PDG2020 lists it: 
\beq
\Delta A_{\bf CP} \equiv  A_{\bf CP}(D^0 \to K^+K^-) - A_{\bf CP}(D^0 \to \pi^+\pi^-) = (- \, 1.54 \pm 0.29 ) \cdot 10^{-3} \; .  
\label{CharmCPV} 
\eeq
It is the first measured {\bf CP} asymmetry in the charm system with more than $5\,\sigma$ uncertainty.  
Previous evidences back in 2012 measured a ten-fold asymmetry difference 
maybe with a $3\,\sigma$ significance. We will discuss that in {\bf Sect.\ref{CPAEXP}} and {\bf Sect.\ref{FUTURE}} with details. 

The weak decays of charm hadrons mostly produce 3- \& 4-body FS.  
In the SM direct {\bf CP} asymmetries cannot happen in favored one, while for SCS produce of ${\cal  O}(10^{-3})$ and basically zero for DCS, 
as we will discuss in {\bf Sect.\ref{CKMMATRIX}}. 
First one can find in 2-body FS, as happen again. However, 3- \& 4-body FS give us a true picture about {\bf CP} asymmetries both with more transitions 
and larger branching rations than for 2-body FS. Non-perturbative QCD has even larger impact there. 
Of course, one needs more refined tools as we will discuss below.

\section{Overview}
\label{OVERVIEW}

Eq.(\ref{OPEGEN}) shows the impact of HQE and its features; actually we know much more, see Eq.(\ref{HEAVYWIDTH}):  
\bea
\nonumber 
\Gamma (H_Q \to f) = \frac{G_F^2 m_Q^5(\mu)}{192\pi^3}|V_{\rm CKM} |^2 \left[ c_3^{(f)} \frac{\langle H_Q |\bar QQ | H_Q \rangle}{2M_{H_Q}}  + 
\frac{c_5^{(f)}}{m_Q^2} \frac{ \langle H_Q |\bar Q i \sigma \cdot G Q | H_Q \rangle}{2M_{H_Q}} \right.
\\
\left. +\sum_i \frac{c^{(f)}_{6,i}}{m_Q^3} \frac{\langle H_Q |(\bar Q\Gamma_i q )(\bar q \Gamma_i Q)| H_Q \rangle}{2M_{H_Q}}
+{\cal O}(1/m_Q^4)\right] _{(\mu)} \; . 
\label{HEAVYWIDTH}
\eea
We will discuss the expectation values of dimension-3, -5 \& -6 here (and previously in Refs.\cite{CICERONE,BSU}), 
namely  $\bar QQ$, $\bar Q i \sigma \cdot G Q$, $\bar Q (i\vec D )^2 Q$, $(\bar Q\Gamma_i q )(\bar q \Gamma_i Q)$ and their connections. 
$\mu^2_{\pi} \equiv  \langle H_Q |\bar Q (i\vec D )^2 Q | H_Q \rangle/ 2 M_{H_Q}$ and  
$\mu^2_G \equiv \langle H_Q |\bar Q i \sigma \cdot G Q | H_Q \rangle/ 2 M_{H_Q}$. Thus $\mu^2_{\pi}$ and $\mu^2_G$ denote the expectation values of the 
"kinetic" and "chromomagnetic" operators. "Sum rules" are ubiquitous tools in many branches of physics that involve sums or integrals over observables such 
as rates and their moments. In this case one talks about "small velocity (SV)" limit in sum rules, where OPE is applicable \cite{SVLIMIT}. Thus there is a rigorous 
lower bound: $\mu_G^2(\omega) \leq \mu^2_{\pi}(\omega)$. To be specific for beauty mesons: $\mu^2_{\pi} \simeq 0.45 \pm 0.1 \; {\rm (GeV)^2}$ and 
$ \mu^2_{G} \simeq 0.35 \pm 0.03 \; {\rm (GeV)^2}$. Furthermore one can describe the impact of $\langle H_Q |(\bar Q\Gamma_i q )(\bar q \Gamma_i Q)| H_Q \rangle$ 
with two terms of $1/m_Q^3$ as {\em D} \& {\em LS} terms named $\rho^3_D$ \& $\rho^3_{LS}$ 
\footnote{$D$ = Darwin term should not confused with the usual word, namely charm mesons $D$; $LS$ = convection current (spin-orbital) term.} 
to discuss the weak decays of heavy mesons. We assume {\bf CPT} invariance giving $\Gamma (H_Q)$ = $\Gamma (\bar H_Q)$.
One can acquire information even from ${\cal O}(1/m_Q^4)$ terms \& some estimates about ${\cal O}(1/m_Q^5)$ ones; however, we will not discuss them in charm 
decays, where one cannot go after accuracy about the impact of HQE.

For $\Lambda_Q=[Q(du)_{j=0}]$ \& $\Xi_Q=[Q(sq)_{j=0}]$ baryons one gets 
\beq
\langle \Lambda_Q |\bar Q i \sigma \cdot G | \Lambda_Q \rangle \simeq 0 \simeq \langle \Xi_Q |\bar Q i \sigma \cdot G | \Xi_Q \rangle \; ; 
\label{LAMBDAQ0}
\eeq
the pair of light quarks carry $j=0$, while one gets $\Omega_Q = [Q(ss)_{j=1}]$ and therefore 
\beq
\mu^2_G (\Omega_Q) \simeq \frac{2}{3} [M^2(\Omega_Q^{(3/2)}) - M^2(\Omega_Q)  ] \neq 0  \; .
\label{OMEGAQNOT0}
\eeq
The situation with charm quarks is more `complex'. Often charm quarks act as heavy quarks, but not all the time; furthermore HQE 
can be applied only in a `qualified way'; i.e., we `paint the landscape' to show the impact of non-perturbative QCD.  
As a first step one looks at diagrams \cite{CICERONE,TIM}: 
\begin{itemize}
\item
"Pauli Interference" ({\bf PI}) diagrams give sizable impact  on $H_Q$ hadrons in general.
\item
$W\; exchange$ \& $W \; annihilation$ diagrams can combine them in one word: {\bf WA}. It gives small impact for $H_Q$ {\em mesons}. 
On the other hand, $W$ exchanges are {\em not} helicity suppressed in the weak decays of charm {\em baryons}; therefore they give large impact there; 
thus we use the word {\em W} Scattering: {\bf WS}.

\end{itemize} 
The best example of heavy quarks is the beauty quark. First one can look at the pattern in the beauty lifetimes; first the pattern \cite{VOLO2000,VARENNA2005}:
\beq
\tau (\Xi_b^0) \simeq \tau (\Lambda_b^0) <   \tau (B^0_s) \simeq  \tau (B^0) <  \tau (\Xi_b^- )  \leq \tau (\Omega_b^-) \; , 
\label{BEAUTY1}
\eeq
where `$<$' means a few \%. PDG2020 tells us: 
\bea
\tau (\Xi_b^0)= (1.480\pm 0.030 )  \cdot 10^{-12}\, {\rm s}&,& \tau (\Lambda_b^0) = (1.471 \pm 0.009 ) \cdot  10^{-12}\, {\rm s}
\\
\tau (B^0_s) = (1.510 \pm 0.004 ) \cdot 10^{-12}\, {\rm s}       &&  \tau (B^0) = (1.519 \pm 0.004 ) \cdot 10^{-12}\, {\rm s} 
\\
\tau (\Xi_b^- ) = ( 1.572 \pm 0.040) \cdot 10^{-12}\, {\rm s}  &,& \tau (\Omega_b^-) = (1.64\, ^{+ 0.18}_{-0.17}  ) \cdot 10^{-12}\, {\rm s}   \; ;  
\label{BEAUTY2}
\eea
it is well described as expected. Compare the ratios, where we best understand the underlying forces including non-perturbative QCD:
\beq
\tau (\Lambda_b^0)/\tau (B^0) |_{\rm PDG2020} \simeq 0.97 \pm 0.01  \; \; {\rm vs.} \; \; \tau (\Lambda_b^0)/\tau (B^0) |_{\rm HQE} \simeq 0.94 \pm 0.04  \; . 
\label{BEAUTY}
\eeq 
On the experiment side the lesson is obvious: it is quite achievement to reach uncertainties of 1\%  with the large backgrounds, see the left side of Eq.(\ref{BEAUTY}). 
However, the situation is much subtle on the theoretical side: 
\begin{itemize}
\item
The weak decays of `free quarks' are described by $[... m_Q^5 (1 + ... \frac{\alpha_S}{\pi}+ ...(\frac{\alpha_S}{\pi})^2 +... )]$

\item
Yet one has to include non-perturbative QCD for bound heavy quarks, namely: 
$[... m_Q^5 (1 + ... \frac{\alpha_S}{\pi}+ ...(\frac{\alpha_S}{\pi})^2 +... ) + ...(\frac{\mu}{m_Q})^2(1+ ...   \frac{\alpha_S}{\pi} +...) +...(\frac{\mu}{m_Q})^3(1+ 
... \frac{\alpha_S}{\pi} +... )...]$. 

\item 
Furthermore, non-perturbative QCD starts at ${\cal O}(1/m_Q)^2$ and continue with ${\cal O}(1/m_Q)^3$ etc. for heavy baryons, while heavy mesons basically start 
with ${\cal O}(1/m_Q)^3$. 

\item
For extreme heavy quarks one gets theoretical limit is not `zero', but close to `one'. 
To be more specific we talk about beauty hadrons. Thus HQE uncertainties in the lifetimes of beauty hadrons is around `one'; i.e., 
a very sizable uncertainty so far, see above in the right side of Eq.(\ref{BEAUTY}). Still theorists who had worked on that can be proud of their achievements. 

\item
Our understanding can be improved based on the semi-leptonic decays of $B^0$ \& $B^0_s$ and $\Lambda_b^0$, $\Xi_b^0$ \& $\Xi_b^-$. 
The connections of these lifetimes and the semi-leptonic decays are not straightway, but crucial. 
\end{itemize}
The main goal of Ref.\cite{CICERONE} and this article is to better understand of {\em charm} dynamics now and for the future, in particular the impact of HQE, 
which is not obvious right away. The `situation' has sizably changed in 2019/20 as we will discuss. Weak transitions of charm hadrons can be applied 
by HQE differently. Previously we had followed the `fashion' about $\tau (\Omega_c^0)$ and $\tau (\Xi_c^0)$.

\section{Experimental tools and 2020 data status}
\label{2020EXP}

Charm physics has been subject of steady interest, in particular since 2003, the date of our Cicerone. 
Many experiments had been active in flavor physics such as CDF, CLEO, BaBar, Belle, BES III. 
A quantum leap in statistics has been provided by LHCb. Mirroring such interest, our community produced several review papers, while the interest of the topic reached 
groups \cite{Amhis:2016xyh} that now maintain regularly the observatory on {\bf CP} violation parameters in the charm system.

\subsection{Recent lifetime measurements of charm hadrons}
\label{LIFEXP}

Up to the  beginning of the second millennium lifetimes of charm hadrons were measured by early fixed-target experiments 
(E687, FOCUS \footnote{It is obvious, why its name was used, although its official name is E831.}, E791, WA89) with typical 50 fs resolutions, and by charm/beauty factories (BaBar, Belle, CLEO-2) at \epem colliders with 150 fs resolutions \cite{Cheung:2001nq}. The former provided the most precise lifetime measurements, the latter contributed with results on time-dependent mixing. The LHCb experiment \footnote{While the results of LHCb are produced by $pp$ collisions, they can be treated 
as fixed-target experiment.} started operation more than ten years ago and has been  producing lifetime result since, 
with very large statistical samples and good control of systematics and backgrounds. 
\par
There are very important differences between the two kind of experiments, reflected on the analysis procedures, the backgrounds, and, therefore, the phenomena which affect systematic errors. Typically fixed-target experiments have excellent 3D vertex and proper time resolutions but large backgrounds, reduced by requiring good vertex separation. Separation between primary and secondary vertices is used as a filter, as well as the reduced proper time variable $t\prime \equiv (L-N\sigma_L)/\beta\gamma c$ which reduces the dependance of proper time on the vertex displacement resolution.
\par
Experiments at \epem  colliders avail of much cleaner environment but can count on poorer time resolution due to the lesser Lorentz boost of charm particles, which result is smaller detachment of secondary vertices. The average  interaction point is normally used for constraining the location of primary vertex. The proper time resolution is often very close (or even larger) of the lifetimes to be measured. For a detailed (although somehow dated) comparative discussion, see Ref.\cite{Cheung:2001nq}.

\begin{figure}[h!]
\begin{center}
\includegraphics[width=22cm]{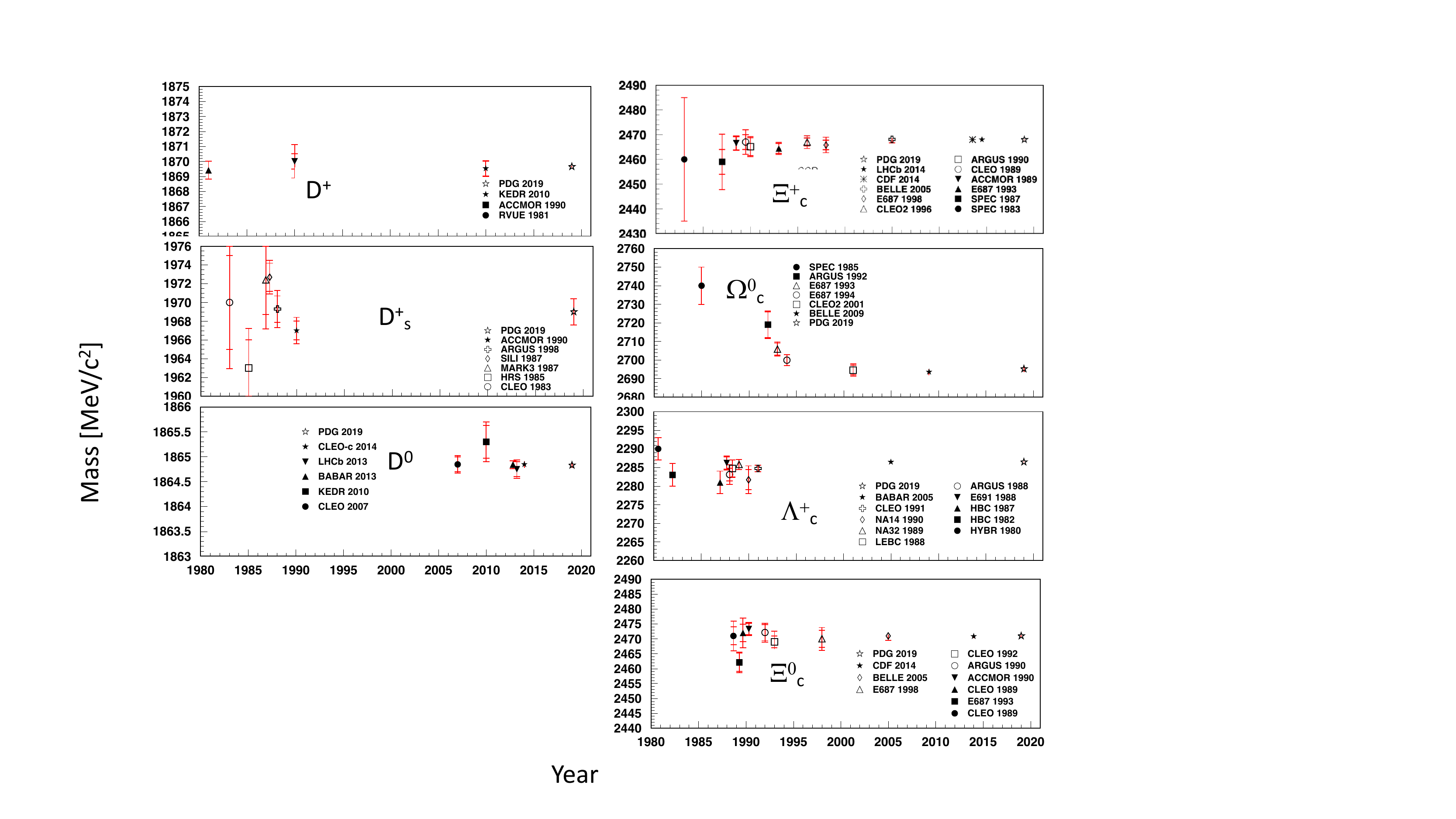}
\end{center}
\vspace{-0.5cm}
\caption{Charm mesons and baryons mass measurements.}
\label{FIG:ALLMASS}
\end{figure}
{\bf Fig.\ref{FIG:ALLMASS}} shows measured masses for charm mesons and  baryons updated to the year 2019; 
one might show the pattern of strong spectroscopy of charm hadrons, namely it is stable since 1985 except the mass of $\Omega_c^0$; 
however, even that was stable from 1995. 

{\bf Fig.\ref{FIG:ALLLIFE}} shows compilations of their weak lifetimes. 
In the past, lifetimes results from fixed-target and \epem experiments have shown inconsistencies and disagreements ({\it tensions} to use a fashionable lingo) 
that sparked discussions on relative strengths and weaknesses of the algorithms used 
\footnote{The lifetime for $\Xi_c^0$ has been enhanced in 2020 sizably what we talk about in the text.}:  
\begin{figure}[h!]
\begin{center}
\includegraphics[width=22cm]{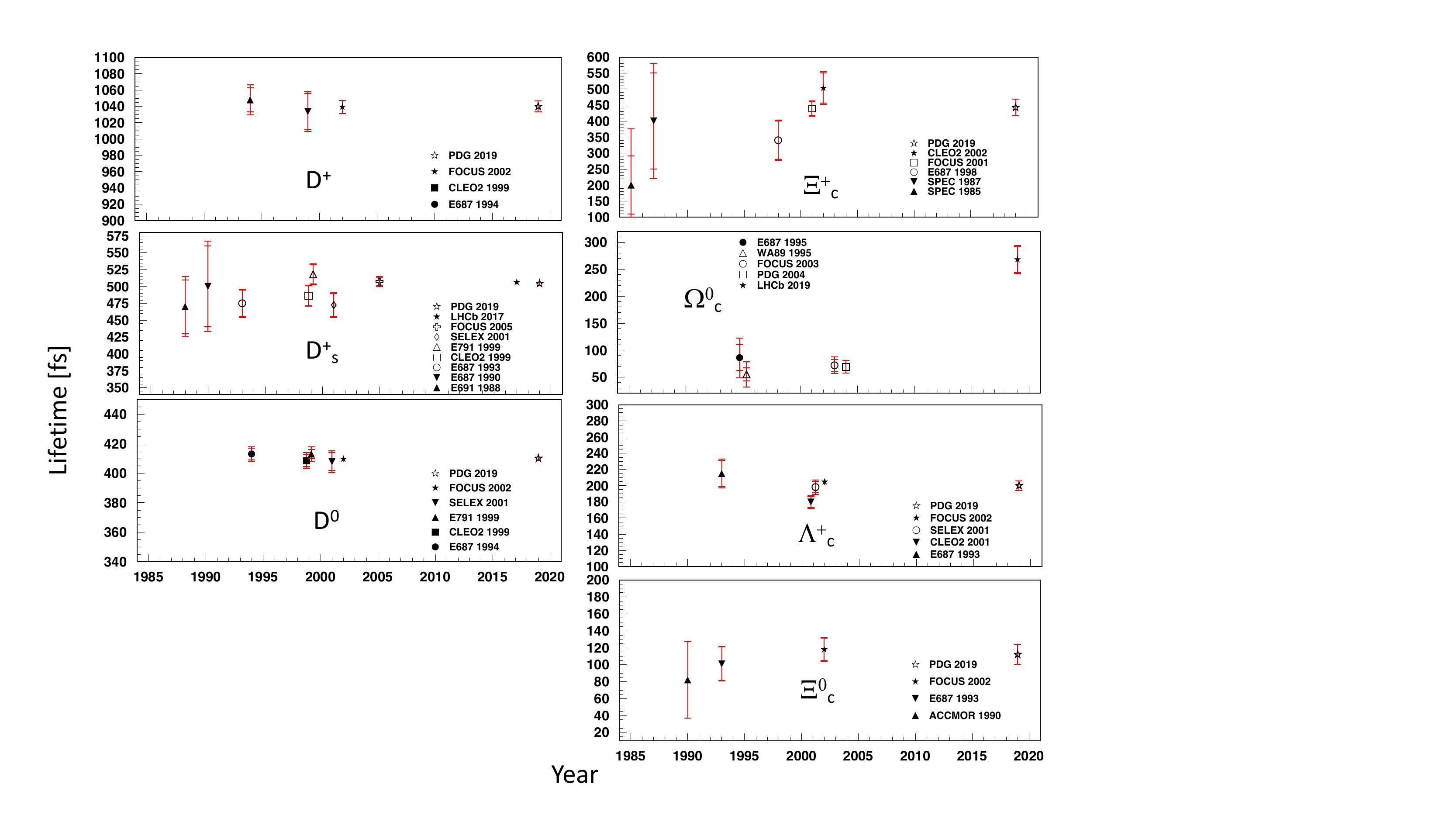}
\end{center}
\vspace{-1.0cm}
\caption{Charm mesons and baryons lifetime measurements.}
\label{FIG:ALLLIFE}
\end{figure}
\par
PDG2018 lifetime averages had shown a hierarchy with charm baryons and mesons:  
\begin{equation}
\tau(\Omega^0_c) < \tau(\Xi^0_c) < \tau(\Lambda^+_c) < \tau(\Xi^+_c) \sim \tau(D^0) < \tau(D_s) < \tau(D^+) \; ; 
\label{LIFE18}
\end{equation}
it is quite different for beauty hadrons (see Eq.(\ref{BEAUTY1})), which is not surprising. We list the lifetimes of $D^+$, $D^+_s$, $D^0$ and 
$\Lambda_c^+$: 
\bea
 \tau(D^+)|_{\rm PDG2018} = (1040 \pm 7) \cdot 10^{-15} \; s  &,& \tau (D^+_s)  |_{\rm PDG2018}= (504 \pm 4) \cdot 10^{-15} \; s
\\
 \tau (D^0)  |_{\rm PDG2018}= (410.1 \pm 1.5)\cdot 10^{-15} \; s  &,&  \tau (\Lambda_c^+ ) |_{\rm PDG2018} = (200 \pm 6) \cdot 10^{-15} \; s
\\
\tau (D^+)/\tau (D^0) |_{\rm PDG2018} \simeq 2.54  & , & \tau (D_s^+)/\tau (D^0) |_{\rm PDG2018} \simeq 1.23
\\
\tau (D^0)/\tau (\Lambda_c^+ ) |_{\rm PDG2018} &\simeq &  2.04
\label{STABLE1}
\eea
This situation is `stable':
\bea
 \tau(D^+)|_{\rm PDG2020} = (1040 \pm 7) \cdot 10^{-15} \; s  &,& \tau (D^+_s)  |_{\rm PDG2020}= (504 \pm 4) \cdot 10^{-15} \; s
\\
 \tau (D^0)  |_{\rm PDG2020}= (410.1 \pm 4) \cdot 10^{-15} \; s  &,&  \tau (\Lambda_c^+ ) |_{\rm PDG2020} = (202.4 \pm 3.1) \cdot 10^{-15} \; s
\\
\tau (D^+)/\tau (D^0) |_{\rm PDG2020} \simeq 2.54  & , & \tau (D_s^+)/\tau (D^0) |_{\rm PDG2020} \simeq 1.23
\\
\tau (D^0)/\tau (\Lambda_c^+ ) |_{\rm PDG2020} &\simeq &  2.03     \; . 
\label{STABLE2}
\eea
However, the landscape of charm baryons is more `complex'. To be specific: 
\bea
\tau(\Omega^0_c)_{\rm PDG2018} &=& (69 \pm 12) \cdot 10^{-15} \; s
\label{OMEGA2018}
\\
\tau (\Xi_c^{0} ) |_{\rm PDG2018} =  (112 ^{+13}_{-10}) \cdot 10^{-15} \; s \; &, &\;  \tau (\Xi_c^{+})  |_{\rm PDG2018}= (442 \pm 26) \cdot 10^{-15} \; s \; .
\eea
It led to
\bea
\tau (\Omega_c^0)/\tau (\Lambda_c^+)  |_{\rm PDG2018} &\sim& 0.35 \pm 0.06
\\
\tau (\Xi_c^0)/\tau (\Lambda_c^+)  |_{\rm PDG2018} \simeq 0.56 \pm 0.06 \; \;  &,& 
\tau (\Xi_c^+)/\tau (\Lambda_c^+)  |_{\rm PDG2018} \simeq 2.21 \pm 0.13
\label{XICH}
\\
\tau (\Xi_c^+)/ \tau (\Xi_c^0) |_{\rm PDG2018} &\sim&  3.9 \pm 0.6      \; .
\label{RATIOBARYON1}
\eea 
In 2019/20 the landscape has changed in qualitative \& quantitative ways:
\beq
\tau(\Xi^0_c) < \tau(\Lambda^+_c) <  \tau(\Omega^0_c)   <  \tau(D^0) <  \tau(\Xi^+_c)  < \tau(D_s) < \tau(D^+) \; . 
\label{LIFE19}
\eeq
The LHCb Collaboration has measured the lifetime of $\tau (\Omega_c^0)$ with a larger value \cite{Omegaclifetime}
\beq
\tau (\Omega_c^0) |_{\rm LHCb, run-1} \simeq (268 \pm 26) \cdot 10^{-15} \, s
\eeq
and $\tau (\Xi_c^0)$, $\tau (\Lambda_c^+) $ \& $\tau (\Xi_c^+)$ with smaller uncertainties \cite{PRECISION}: 
\bea
\tau (\Xi_c^0) |_{\rm LHCb, run-1} &\simeq& (154.5 \pm 2.5) \cdot 10^{-15} \,  s
\\
\tau (\Lambda_c^+)  |_{\rm LHCb, run-1} &\simeq& (203.5 \pm 2.2)\cdot 10^{-15} \, s
\\
\tau (\Xi_c^+)  |_{\rm LHCb,  run-1} &\simeq& (456.8 \pm 5.5)\cdot 10^{-15} \, s    \; . 
\eea
The LHCb measurement of $\Omega^0_c$ lifetime is based on a collected  final sample five times larger than those accumulated by all predecessors, 
and yields a  lifetime four times larger, much beyond the errors. The LHCb measurement is performed on a sample of b-tagged $\Omega^0_c$ decays. 
The variable used is proper time rather than the $t ^\prime$ reduced proper time. The proper time resolution declared is 
$(80 - 100)\cdot 10^{-15}\, s$. 
To reduce the systematics, LHCb normalize to  $D^+$ decays. 
The LHCb measurement of $\Omega^0_c$ lifetime is very relevant:  it changes the hierarchy of lifetimes settled until 2018, and it has stemmed a vibrant discussion in the community. The lifetime value measured is so large that could have been easily measured much earlier than 2018  by experiments at \epem whose resolution is about 
$150 \cdot 10^{-15} \, s$ typically  \cite{Cheung:2001nq}. 
CLEO-c and Belle have both observed $\Omega^0_c$ and measured its mass, they should/could have measured quite easily the lifetime value measured by LHCb.
As an example and comparison: Belle has measured the lifetime of the $\tau$ lepton with $(290.17 \pm 0.53 \pm 0.33) \cdot 10^{-15}\, s$. 

It has convinced the PDG team to use their values 
\bea
\tau(D^+) |_{\rm PDG2020} = (1040 \pm 7 ) \cdot 10^{-15} \, s            &,&  \tau(D^+_s)|_{\rm PDG2020} = (504 \pm 4 ) \cdot 10^{-15} \, s
\\
\tau (\Xi_c^+)|_{\rm PDG2020} &=& (456 \pm 5 ) \cdot 10^{-15} \, s  
\\
\tau(D^0)|_{\rm PDG2020} = (410.1 \pm 1.5 ) \cdot 10^{-15} \, s          &,&  \tau( \Omega^0_c)|_{\rm PDG2020} = (268 \pm 26 ) \cdot 10^{-15} \, s
\\
\tau(\Lambda^+_c)|_{\rm PDG2020} = (202.4 \pm 3.1 ) \cdot 10^{-15} \, s &,& \tau (\Xi_c^0)|_{\rm PDG2020}= (153 \pm 6 ) \cdot 10^{-15} \, s  
\label{LIFE20}
\eea
or these ratios
\bea
\tau (\Omega_c^0)/\tau (\Lambda_c^+)  |_{\rm PDG2020} &\simeq& 1.32 \pm 0.10
\label{RATIOOMEGACH}
\\
\tau (\Xi_c^0)/\tau (\Lambda_c^+)  |_{\rm PDG2020} \simeq 0.76 \pm 0.05 \; \;  &,& 
\tau (\Xi_c^+)/\tau (\Lambda_c^+)  |_{\rm PDG2020} \simeq 2.25 \pm 0.05
\label{RATIOXICH}
\\
\tau (\Xi_c^+)/ \tau (\Xi_c^0) |_{\rm PDG2020} &\sim&  2.98 \pm 0.05  \; . 
\label{RATIOXICHXICH}
\eea 
One can `paint' the landscape of lifetimes of charm hadrons; it is not $\tau (D^+)/\tau (\Omega_c^0)|_{\rm data}$ $\sim 13$ in PDG2018, 
but $\tau (D^+)/\tau (\Xi_c^0) |_{\rm data} \sim 7$ with a different `actor' in PDG2020.

\subsection{Semi-leptonic decays}

One can look at the measured ratios of semi-leptonic charm mesons: 
\bea
{\rm BR}(D^+ \to e^+ X)|_{\rm PDG2020} &=& (16.07 \pm 0.30)\% 
\\ 
{\rm BR}(D^0 \to e^+ X)|_{\rm PDG2020} &=& (6.49 \pm 0.11)\%
\\
{\rm BR}(D^+_s \to e^+ X) |_{\rm PDG2020} &=& (6.5 \pm 0.4)\%  \; . 
\label{SLD}
\eea 
These values have hardly changed over time. 

The situations are quite different for charm baryons, namely the semi-leptonic width has been measured only for $\Lambda_c^+$: 
\beq
{\rm BR}(\Lambda_c^+ \to e^+ X)|_{\rm PDG2020} = (3.95 \pm 0.35)\%  \; . 
\eeq
We mention 
\beq
\frac{{\rm BR}(\Lambda_c^+ \to e^+ X)}{{\rm BR}(D^0 \to e^+ X)}  |_{\rm PDG2020} \simeq 0.61 \pm 0.05 \; \; \; \; \;  {\rm vs.} \; \; \; \; \; 
\frac{ \tau (\Lambda_c^+)}{\tau (D^0) }|_{\rm PDG2020} \simeq 0.49 \pm 0.01 \; ; 
\eeq
the situations with charm baryons are more `complex', as we will discuss in {\bf Sect.\ref{SEMILEPTON}}.

\subsection{Recent {\bf CP} asymmetry measurements}
\label{CPAEXP}

The formalism of {\bf CP} asymmetries is described in Ref.\cite{CICERONE} as well as in numerous other articles (see \cite{Polycarpo:2014saa} for a much detailed review of 
mixing and {\bf CP} violation). 

There are three classes, although there is only one member that can contribute to all, namely $D^0$, while all charm hadrons can contribute to the third one:
\begin{itemize}
\item
It needs $D^0 - \bar D^0$ oscillations with only $\Delta C=2$ transitions. It has at least been established due to $y_D \neq 0$. 
The cleanest way to probe {\bf CP} violation is to compare 
$D^0 \to l^-X$ vs. $\bar D^0 \to l^+ \bar X$. It means $|q/p|_D \neq 1$, yet it is time {\em independent}. PDG2020 tells us:
\beq
1- |q/p |_D = 0.08 \, ^{+0.12}_{-0.09}  \; ;
\eeq
it is unlikely to establish $|q/p |_D \neq 1$ `soon'.

\item 
One can discuss weak decays of $D^0$ to {\bf CP} eigenstates like $f=K^+K^-$ and/or $f=\pi^+\pi^-$. 
Thus {\bf CP} asymmetries can be described by Im$[(\frac{q}{p})_D\frac{\bar A_f}{A_f}]$; it shows the   
interplay of $\Delta C=2$ with  $\Delta C=1$ ones. One can probe {\bf CP} asymmetries with time-{\em dependent} decay rates:
\beq
A_{\bf CP} (f; t) \equiv \frac{\Gamma (D^0 (t) \to f) - \Gamma  (\bar D^0 (t) \to f)}{\Gamma (D^0 (t) \to f) + \Gamma  (\bar D^0 (t) \to f)} \; . 
\label{TIMDEP}  
\eeq 
$D^0 - \bar D^0$ oscillations are described by $x_D  \equiv \Delta M_D/\Gamma_D$ and $y_D \equiv \Delta \Gamma_D/ 2\Gamma_D$. 
We know that are small: $x_D, y_D \sim {\cal O} (10^{-3})$; 
to make it clearer: $y_D \sim (0.64 \pm 0.09) \cdot 10^{-2}$ and $x_D \sim (0.32 \pm 0.14 ) \cdot 10^{-2}$.
Thus  \footnote{While these Equations have been used one way or other, to our knowledge Tommaso Pajero has shown that in details for the first time \cite{MICHAEL}.}
\bea
A_{\bf CP} (f; t) &\simeq & A^{\rm dir}_{\bf CP} (f) - \frac{t}{\tau_{D^0}} A_{\Gamma} (f)
\label{ADIR}
\\
A_{\Gamma} (f) &\simeq &- x_D \,\phi_f + y_D ( |q/p|_D -1) - y_D\, A^{\rm dir}_{\bf CP} \; .
\label{AGAMMA}
\eea
There are two observables that probe {\bf CP} violation: $A^{\rm dir}_{\bf CP} (f)$ and $A_{\Gamma} (f)$; 
they can be differentiated by their time {\em dependences}. The amplitudes $\bar A(D^0 \to f)$ \& $A(D^0 \to f)$ can be described by $\Delta C=1$ dynamics only. 
A time-integrated {\bf CP} asymmetry can be measured. 
To first order of $D^0 - \bar D^0$ oscillation one can describe it as 
\beq
\langle A_{\bf CP} (f; t) \rangle \simeq A^{\rm dir}_{\bf CP} (f) - \frac{\langle t(f) \rangle}{\tau _{D^0}}   A_{\Gamma} (f)   \; . 
\label{FIRST}
\eeq
Assuming $\phi_f \simeq \phi$ = arg$(q/p)$ it is fine for now \cite{LHCbNOVDh+h-} \item
One can learn from the histories of {\bf CP} asymmetries in $D^0 \to h^+h^-$.  
The first set of results on $A_{\bf CP} (K^-K^+)$ and $A_{\bf CP} (\pi^- \pi^+)$ is customarily attributed to the Fermilab fixed target experiments E791 and FOCUS. 
Both measured zero $A_{\bf CP}$ asymmetries with uncertainties of $\sim$ 13 \% \cite{Aitala97, Aitala98} and 6\% \cite{Link:2000aw}, respectively. 
Both E791 and FOCUS only quoted asymmetries without quoting asymmetry differences. Neither results, therefore, are recorded in PDG2020. 

\end{itemize}
Only after nearly a decade the 1\% precision level was attained at \epem colliders by Belle \cite{Staric:2008rx} 
--  $A_{\bf CP}(D^0 \to K^+K^-)= (-\, 0.43 \pm 0.30 \pm 0.11) \cdot 10^{-2}$ and $A_{\bf CP}(D^0 \to \pi^+\pi^-)= (+\, 0.43 \pm 0.52 \pm 0.12) \cdot 10^{-2}$  -- 
and BaBar  \cite{Aubert:2008yd}: $A_{\bf CP}(D^0 \to K^+K^-\pi^0)$ and $A_{\bf CP}(D^0 \to \pi^+\pi^-\pi^0)$ difference from zero with uncertainties of $10^{-2}$ or less. 

Big excitement arose in 2012 when LHCb at the CERN LHCb collected a $10^6$ decay sample and measured: 
$\Delta A_{\bf CP}= A_{\bf CP} (D^0 \to K^+K^-) - A_{\bf CP} (D^0 \to \pi^+\pi^-)  = (-\, 0.82 \pm 0.21\pm 0.11) \cdot 10^{-2}$ \cite{LHCb:2012}; 
{\it i.e.,} a result deviated from zero with a significance of $3.5\,\sigma$ just enough to claim an evidence. 

CDF at the Fermilab Tevatron soon thereafter - and with a similar size sample - measured 
$\Delta A_{\bf CP}= (-\, 0.46 \pm 0.31 \pm 0.12) \cdot 10^{-2}$ \cite{Aaltonen:2011se}, 
and immediately later $\Delta A_{\bf CP}=(- \, 0.62\pm0.21\pm0.10) \cdot 10^{-2}$ \cite{CDF:2012qw}, 
result compatible with LHCb but with a significance even lower, $2.7 \,\sigma $ difference from zero. 
In the  Summer of 2012, Belle presented a result that remained preliminary: $\Delta A_{\bf CP}=(-\, 0.87\pm 0.41\pm 0.06) \cdot 10^{-2}$ \cite{Ko:2012px}, 
basically confirming both LHCb and CDF, with a very low significance of $2.1 \,\sigma $. 
These results suggesting {\bf CP} violation in the ballpark of 1\% immediately stimulated a flurry of theory work.

However, over the years 2012 -- 2016 LHCb accumulated event samples of ${\cal O}(10^7)$ presenting results 
compatible with zero asymmetry, and errors slowing decreasing down to 0.1\%, see Refs.\cite{Aaij:2013bra, Aaij:2013ria, Aaij:2014gsa}. 
The $\Delta A_{\bf CP}$ saga seemed to turn to a focal point in 2019 with LHCb precise measurement \cite{LHCbGuy} based on $7 \cdot 10^7$ event data sample: 
$\Delta A_{\bf CP}(D^0 \to K^+K^-/\pi^+\pi^- )=(-\,  0.154 \pm 0.029) \% $, see above. 
A synopsis of $\Delta A_{\bf CP} $ measurements is shown in {\bf Fig.\ref{FIG:HISTORYCPASYM}}. 

The next step is to probe indirect {\bf CP} violation in the decays of $D^0$ in the LHCb data including the total result from run-1 as 3 fb$^{-1}$ and 
from run-2 as 5.4 fb$^{-1}$ \cite{LHCbNOVDh+h-}: 
\bea
A_{\Gamma} (K^+K^-) &=& - \, (4.4 \pm 2.3 \pm 0.6) \cdot 10^{-4}
\\
A_{\Gamma} (\pi^+ \pi^-) &=& + \, (2.5 \pm 4.3 \pm 0.7) \cdot 10^{-4}  \; . 
\eea
Our community knows that re-scattering due to QCD happens for two reasons, 
in particular in the region of 
energies for charm hadrons as discussed in Ref.\cite{CICERONE}.with some details: non-perturbative QCD leads to re-scattering 
of hadrons as $\bar KK \leftrightarrow \pi \pi$. There are two statements: (a) When one looks at the systematics uncertainties, one has to wait for run-3 
of LHCb and/or Belle II. 
(b) One has to probe 3- \& 4-body FS, see  {\bf Sect.\ref{RESC}} \footnote{Item (b) is not favored by experimenters.}.  
We had pointed out before \cite{CICERONE}. We will come back to that in {\bf Sect.\ref{CPVSCS}} \footnote{The history of  {\bf CP} asymmetry in the transitions of charm mesons is very different  than for strange and beauty mesons.}.  

Actually, there is a third statement, which is quite different from systematics uncertainties, namely to use $D^0 \to K^+K^-$ \& $D^0 \to \pi^+\pi^-$ 
from  $D^* (2010)^+ \to D^0 \pi^+$. The present analysis corresponds to 1.9 fb$^{-1}$ by the LHCb detector at 13 TeV. It has shown 
$A_{\Gamma}(K^+K^-) = +\, (1.3 \pm 3.5 \pm 0.7 ) \cdot 10^{-4}$ and $A_{\Gamma}(\pi^+ \pi^-) = +\, (11.3 \pm 6.9 \pm 0.8 ) \cdot 10^{-4}$ \cite{MICHAEL}; 
it will continue.

\section{Theoretical tools}
\label{PHENO}

Also the box of theoretical tools has been improved this century in different directions.  

\subsection{CKM matrix}
\label{CKMMATRIX}

The Wolfenstein representation of the CKM matrix had described the landscape with three families in 1983 \cite{WOLFMAT}; it is very usable and mostly used  
(including our 2003 `Cicerone') \footnote{As we had said in our `Cicerone':  the SM is incomplete.}:
\bea
\lambda =  0.22453 \pm 0.00044      &,& A = 0.836 \pm 0.015
\\
\bar \rho =  0.122 ^{+0.018}_{-0.017}     &,& \bar \eta = 0.355 ^{+0.012}_{-0.011}  \; . 
\label{WOLF}
\eea  
The value of the Cabbibo angle is very well measured, while the other three parameters should be of the ${\cal O}(1)$; indeed, the value of $A$ is fine. 
However, to fit the data with the other two parameters are {\em not}, see the second line above. In the world of quarks no sign of ND 
has been established yet. 

Therefor our community has to go after accuracy or even precision; thus one has to use a consistent parametrization of the 
CKM matrix. The best example so far is described in Ref.\cite{AHN}. In addition to $\lambda$ it gives two angles plus one phase:
\bea
f = 0.754 ^{+0.016}_{- 0.011}  \; &, &\; \bar h = 1.347 ^{+0.045}_{-0.030}
\\
\delta_{\rm QM} & = & (90.4^{+0.36}_{-1.15} )^o
\label{AHN}
\eea
There is a special case: the SM gives basically zero {\bf CP} asymmetries in DCS transitions. For the results of the LHCb experiment one has to wait for run-3 data 
and for real results from Belle II during the 2020's; i.e., DCS is a hunting region for ND! Tiny rates are not the only challenge: experimental uncertainties could give 
Cabibbo favored transition $\Lambda_c^+ \to p K^-\pi^+$ `seen' as DCS $\Lambda_c^+ \to p K^+\pi^-$.

\subsection{Re-scattering}
\label{RESC}

We had discussed the impact of (strong) re-scattering in details, see Section 11 `CP violation' in Ref.\cite{CICERONE}; 
thus we add a few comments to update the present situations. Re-scattering 
gives impact on 2-body FS like $\pi^+ \pi^- \leftrightarrow \pi^0 \pi^0$, $K^+K^- \leftrightarrow \bar K^0 K^0$, $\pi \pi \leftrightarrow \bar K K$ \& 
$\pi K \leftrightarrow K \pi$  with non-perturbative QCD in the region $\sim$ 1 - 2 GeV \footnote{One can add $\eta$ \& $\eta^{\prime}$.}. 
However, 2-body FS of non-leptonic weak decays are a small part of charm hadrons (\& tiny ones for beauty hadrons). 
It means one need much more information about the underlying dynamics and re-fined tools. It is crucial to analyze re--scattering 
$ 2 \to 3, 4, ...$ for FS.  There is a price for working on 3- \& 4-body FS, but also a prize for the underlying dynamics, namely the existence of ND \& its features. 
There is also a good sign: for charm hadrons the FS hardly go beyond 4-body FS. We can go beyond general statements, 
namely to describe the amplitude of an initial state to the final one; first we look  at simple case, where the FS of two classes, namely 
$a$ \& $b$ amplitudes (like for non-leptonic decays of $K^0$):
\bea
T(P \to a) = e^{i\delta_a} ( T_a +  T_{b} \, i \, T^{\rm resc}_{b a} ) \; &, & \; 
T(\bar P \to \bar a) =  e^{i\delta_a}  ( T_a^* + T^*_b \, i \,  T^{\rm resc}_{b a}) 
\\
\Gamma (P) = \Gamma (P \to  a) + \Gamma ( P \to b) &=& \Gamma (\bar P \to \bar a) + \Gamma (\bar P \to \bar b) = \Gamma (\bar P)
\eea
Thus 
\bea
\Delta \Gamma (a) \equiv  |T(\bar P \to \bar a)|^2 - |T(P \to a)|^2 = 4 \, T^{\rm resc}_{a b}\;  {\rm Im} \,T_a^* T_b  
\\
\Delta \Gamma (b) \equiv |T(\bar P \to \bar b)|^2 - |T(P \to b)|^2 = 4 \, T^{\rm resc}_{b a}\;  {\rm Im} \,T_b^* T_a
\eea
with $T^{\rm resc}_{a b}$ = $T^{\rm resc}_{b a}$ = $(T^{\rm resc}_{a b})^*$. 
Therefore
\beq
\Delta \Gamma (a) = - \Delta \Gamma (b)
\eeq
as expected: re-scattering is based on QCD (\& QED) and {\bf CPT} invariance is assumed. 

This simple scenario is easily be extended to two sets of $A$ and $B$ of FS:  all states $a$ in set $A$ the transition amplitudes have the 
same weak couplings and likewise for states $b$ in set $B$. One then finds due to {\bf CPT} invariance: 
\bea
\Delta \Gamma (a) = 4 \, \sum_{c\; \epsilon \; A} T^{\rm resc}_{a c}\;  {\rm Im} \,T_a^* T_c &=& - \; 4\,  \sum_{d\; \epsilon \; B} T^{\rm resc}_{d b}\;  {\rm Im} \,T_b^* T_d
= -\;  \Delta \Gamma (b) 
\label{MANYBODY}
\\
\sum_a \Delta \Gamma (a) &=& 4 \sum_a \sum_{a\neq c} T^{\rm resc}_{a c}\;  {\rm Im} \,T_a^* T_c  = 0 \; , 
\eea
where $T^{\rm resc}_{a c}$ \& ${\rm Im} \,T_a^* T_c $ are symmetric \& anti-symmetric, respectively. 

Our community has a good report about probing 3-body FS with Dalitz plots, we will discuss just next and made some progress about 4-body FS. 
To understand more information from the data, one needs several tools like chiral symmetry and dispersion relations. 
Dalitz plots with $\pi$, $K$, $\eta$ \& $\eta^{\prime}$ probe the underlying dynamics with two observables (for charm mesons): 
without angular correlations a plot is flat, while resonances and thresholds show their impact. One expects that also broad resonances in the 
0.5 - 1.5 GeV; scalar ones like $f_0(500)/\sigma$, $K^*_0(700)/\kappa$ etc. etc. should have impact; one should remember that these ones can{\em not} be described 
with Breit-Wigner parameterizations.

\subsection{Probing {\bf CP} asymmetries in 3- \& 4-body FS}
\label{DALITZ}

Dalitz plots have been suggested to probe parity violation \cite{DALITZ}. Our community had continued to probe {\bf CP} asymmetries 
in the transitions of beauty hadrons, which were found. The next step is to apply it to 3-body FS charm hadrons. 
One goes after {\bf CP} violation with{\em out} production asymmetries. 
Next one probes {\bf CP} asymmetries 
with four-body FS in more subtle ways; one example from beauty baryons: $\Lambda_b^0 \to p\pi^-\pi^+\pi^-$ gives the first evidence of 
{\bf CP} asymmetry from run-1 \cite{LHCBBARCP}. One of the first LHCb paper from run-2 are very interesting \cite{LHCBPV}: 
(a) {\bf P} asymmetry has been established in $\Lambda_b^0 \to p \pi^-\pi^+\pi^-$; still we cannot understand the lesson we have learnt from the data. 
(b) {\bf CP} asymmetry has not been established in beauty baryons; we have to wait for run-3 results -- or somebody will find another `road'  
about {\bf CP} asymmetry in baryons based in run-2 data. .

\subsection{"Duality" between hadrons and quarks}
\label{DUAL}

Now we come back to re-scattering in a different way. In general, "duality" is not `local'; i.e., one has to use averaged one 
over  the region $\sim$ 1 - 1.5 GeV. On the other hand, when one is {\em not} close to thresholds \& resonances, one can use `local duality'  
including perturbative QCD. We will come back below.

\section{Non- \& semi-leptonic widths of charm hadrons}
\label{WIDTHHADRONS}

The ground states of charm mesons (baryons) are 3 (4) ones that decay only weakly.  

\subsection{Lifetimes of $D^0$, $D^+$ \& $D^+_s$ and their semi-leptonic widths}
\label{MESONS}

In {\bf Sect.\ref{2020EXP}} we have listed the data from PDG2020. Now one can compare the SM predictions from HQE mostly due to {\bf PI}, 
but some impact of {\bf WA} \cite{CICERONE} based on 
$\mu_{\pi}|_{D}\sim 0.45  \; (\rm GeV)^2$ [vs.  $\mu_{\pi}|_{B}\simeq 0.37 \; (\rm GeV)^2$]  
and $\mu_G|_{D}\sim 0.41  \; (\rm GeV)^2$ [vs.  $\mu_G|_{B}\simeq 0.37 \; (\rm GeV)^2$]: 
\bea
\tau (D^+)/\tau (D^0)|_{\rm HQE} \sim 2.4 \; &,& \; \tau (D^+_s)/\tau (D^0)|_{\rm HQE} \sim 0.9 - 1.3
\\
\left. \frac{{\rm BR}(D^+ \to e^+\nu X)}{{\rm BR}(D^0 \to e^+\nu X)} \right|_{\rm HQE} \sim 2 \;  &,& \; 
\left. \frac{{\rm BR}(D^+_s \to e^+\nu X)}{{\rm BR}(D^0 \to e^+\nu X)}  \right|_{\rm HQE} \sim 1.2
\label{HQEm_c} 
\eea 
One can compare with the present data: 
\bea
\tau (D^+) /\tau (D^0) |_{\rm PDG2020} \simeq 2.5  \; \; &,& \; \; \tau (D^+_s) /\tau (D^0)|_{\rm PDG2020} \simeq 1.2 
\\
\left. \frac{{\rm BR}(D^+ \to l^+\nu X)}{ {\rm BR} (D^0 \to l^+ \nu X)} \right|_{\rm PDG2020}  \simeq  2.4  \; \; &,& \; \; \left. 
\frac{{\rm BR}(D^+_s \to l^+\nu X)}{ {\rm BR} (D^0 \to l^+ \nu X)} \right|_{\rm PDG2020}  \simeq 1
\label{PDG2020}
\eea
HQE is successful already semi-quantitatively for charm mesons due to its impact of ${\cal O}(1/m_c^3)$, 
Actually it is amazing with $\mu \sim$ 1 GeV \& $m_c \sim 1.3$ GeV  
\footnote{Uraltsev had pointed out that we have stronger control over semi-leptonic than for non-leptonic ones.}. 
As discussed in Ref.\cite{CICERONE} with details, it is {\em not} a miracle.

More recent analysis has given \cite{LENZ}:
\beq
\tau (D^+)/\tau(D^0)|_{\rm HQE2013} \simeq 2.2 \pm 0.4 \; , \; \tau (D^+_s)/\tau(D^0)|_{\rm HQE2013} \simeq 1.19 \pm 0.12  \; . 
\eeq  
We are not convinced (yet) that the uncertainties are so small; one point is we have little control over ${\cal O}(1/m_c^4)$ contributions. 

\subsection{Lifetimes of $\Lambda_c^+$, $\Xi_c^{+}$, $\Xi_c^{0}$ \& $\Omega_c^0$}
\label{BARYONSLIFE}

There are four charm baryons that decay only weakly with a single charm quark: two carry isospin zero 
($\Lambda_c^+$ \& $\Omega_c^0$), while the other two are isospin 1/2 ($\Xi_c^{+}$ \& $\Xi_c^{0}$). 
First one looks at the PDG data from 2002, when our "Cicerone" was produced, or PDG2018. The values of lifetimes of $\Lambda_c^+$ \& 
$\Xi_c^+$ have not changed from 2002 to 2018, while the ones for $\Xi_c^0$ \& $\Omega_c^0$ consistent with them. 

On fairly general grounds a hierarchy had been predicted \cite{VOLOSHIF86,BBD} 
\footnote{To be honest: our 2003 "Cicerone" had followed the previous literature.}:
\beq
\tau ( \Omega_c^0) < \tau (\Xi_c^{0} ) < \tau (\Lambda_c^+ ) < \tau (\Xi_c^{+})  \; . 
\label{CHARMBAR?}
\eeq
These analyses invoked the assumption that a valence quark description provides a good approximation of these rates. 
If these hadrons contained large `sea' components, they would all share the same basic reactions, albeit in somewhat 
different mixtures. We will come back to that below to deal with the 2019 situations with somewhat surprising results. 

It had been suggested to describe qualitatively the patterns in the measured lifetimes of charm and strange baryons, see PDG2018 
\footnote{These baryons carry spin 1/2 with one exception: $\Omega^-$ carries spin 3/2; it was one of several reasons, why one needs 
"color" quantum number.}: 
\bea
\tau (\Omega_c^0) < \tau (\Xi_c^0) &<& \tau (\Lambda_c^+) <   \tau (\Xi_c^+) 
\\
\tau (\Omega^-) < \tau (\Xi^-) &<& \tau (\Lambda^0) <   \tau (\Xi^0)  \; . 
\label{STRANGE}
\eea
$\Delta$(charge) = -1 is a filter for the connection for these baryons or the quarks $c \leftrightarrow s$. 
These patterns had been seen as `natural' for a long time (for theorists). One can describe the situation including spectroscopy: 
(a) It connects two pairs of isospin singlets, namely $\Omega_c^0 = [css]$ with $\Omega^- = [sss]$ and $\Lambda_c^+=[cud]$ with 
$\Lambda^0 = [sud]$. (b) It connects isospin doublets $\Xi_c^0 = [csd]$ with $\Xi^- = [ssd]$ and $\Xi_c^+ = [csu]$ with $\Xi^0 = [ssu]$; 
i.e., it changes due to re-scattering of $d \to u$ \footnote{For this argument it does not matter, if $s$ quarks are "current" or `constituent' ones.}. 

Through the year of 2018 it had been viewed as a `stable' situation \footnote{It included us.}. 
One could see an analogy with strange baryons, although on very different scales. 
However, one author did not agree with such a statement \cite{HYC2018}:
\beq
230 \cdot 10^{-15} \; {\rm s} \; < \; \tau (\Omega_c^0  )  \; < \;  330 \cdot 10^{-15} \; {\rm s}  \; ;
\label{HYC}
\eeq
it had shown good `judgment'. 

The landscape has sizably changed in 2019 --   $\tau ( \Omega_c^0)$ -- and 2020 -- $\tau (\Xi_c^{0})$:
\bea
\nonumber
\tau (\Lambda_c^+ ) |_{\rm PDG2020} = (202.4 \pm 3.1) \cdot 10^{-15} \; s \; &,& \;  \tau ( \Omega_c^0) |_{\rm PDG2020} = (268 \pm 26) \cdot 10^{-15} \; s
\\
\nonumber
\tau (\Xi_c^{0} ) |_{\rm PDG2020} =  (153 \pm 6) \cdot 10^{-15} \; s     &,& \tau (\Xi_c^{+})  |_{\rm PDG2020} = (456 \pm 5) \cdot 10^{-15} \; s \; .  
\eea
First one looks at the 2019/20 pattern; 
\beq
\tau (\Xi_c^0) < \tau (\Lambda_c^+) < \tau (\Omega_c^0) < \tau (\Xi_c^+) \; .  
\eeq
Now we have learnt that even the pattern of measured lifetimes of charm baryons is different from strange baryons; this `challenge' has disappeared.  

One can compare the ratios of the lifetimes of charm baryons from PDG2019/20,   
see Eq.(\ref{RATIOOMEGACH}), Eq.(\ref{RATIOXICH}) \& Eq.(\ref{RATIOXICHXICH}), with expectations based on HQE; 
the landscape is even more complex for charm baryons. 
\bea
\tau (\Xi_c^0)/\tau (\Lambda_c^+)  |_{\rm PDG2020} \sim 0.76  \; \;  &,& \; \tau (\Xi_c^0)/\tau (\Lambda_c^+)  |_{\rm HQE} \sim  0.5
\\
\tau (\Xi_c^+)/\tau (\Lambda_c^+)  |_{\rm PDG2020} \sim 2.25  \; &,& \; \tau (\Xi_c^+)/\tau (\Lambda_c^+)  |_{\rm HQE} \sim 2.2 
\\
\tau (\Xi_c^+)/ \tau (\Xi_c^0) |_{\rm PDG2020} \sim  3  \; &,& \; \tau (\Xi_c^+)/ \tau (\Xi_c^0) |_{\rm HQE} \sim  2.8  \; . 
\eea
HQE gives qualified predictions, when one is `realistic'. Up to 2018 one can see there is an exception: 
$\tau (\Omega_c^0)/\tau (\Lambda_c^+)  |_{\rm PDG2020} \sim 1.34$ vs. $ \tau (\Omega_c^0)/\tau (\Lambda_c^+)  |_{\rm `naive' \; HQE} \sim 0.4$. 
However, one has to apply `sensible' HQE \cite{HYC2018}: 
\beq
 \tau (\Omega_c^0)/\tau (\Lambda_c^+) |_{\rm PDG2020} \sim 1.34 \; , \;  \tau (\Omega_c^0)/\tau (\Lambda_c^+)  |_{\rm `sensible' \; HQE} \sim 1.4 \; . 
\eeq 

HQE basically applies first 
to $D^0$ \& $D^+_{(s)}$ decays as ${\cal O}(1/m_c^3)$ terms with semi-quantitative successes as discussed above. Yet it applies already for charm baryons 
as ${\cal O}(1/m_c^2)$ terms, but still gives large contributions qualitatively as ${\cal O}(1/m_c^3)$ ones. 
There are four baryons with $C=1$ that decay weakly: $\Lambda_c^+ = [c(ud)_{j=0}]$, $\Omega_c^0 = [c(ss)_{j=1}]$ with $I(J^P)= 0(\frac{1}{2}^+ )$ and 
$\Xi_c^+=[c(su)_{j=0}]$ \& $\Xi_c^0=[c(sd)_{j=0}]$ with $I(J^P)= \frac{1}{2}(\frac{1}{2}^+ )$.  
The $J^P$ have not been measured for $\Xi_c^+$, $\Xi_c^0$ \& $\Omega_c^0$; $\frac{1}{2}^+$ is assumed for the quark {\em model} predict; 
the small $j$ is the spin of the two {\em light} quarks $q=u,d,s$. 
One has to re-think about previous assumptions \footnote{Or to use different words: `harbinger'/`augury'.}. 

The `weak' side of HQE applying to charm {\em baryons} is to use quark {\em models}, {\em not} QCD. 
There are more points to compare data with HQE expectations. 
\begin{itemize}
\item
As discussed in details in Ref.\cite{CICERONE} -- see also in Eq.(\ref{LAMBDAQ0}) \& Eq.(\ref{OMEGAQNOT0}) here -- one uses 
$\mu^2_{\pi} \equiv \langle H_Q |\bar Q(i\vec D )^2  Q | H_Q \rangle /2M_{H_Q}$ \&  
$\mu^2_G \equiv \langle H_Q |\bar Q \frac{i}{2}\sigma \cdot G  Q | H_Q \rangle /2M_{H_Q}$ and 
$\mu^2_{\pi}(H_Q)  \geq  \mu^2_G (H_Q)$ are known from QCD in general. 
Furthermore $\mu ^2_G (\Lambda_Q,  \, 1 \, {\rm GeV}) \simeq 0 \simeq  \mu ^2_G (\Xi_Q,  \, 1 \, {\rm GeV})$, while 
$\mu^2_G (\Omega_Q,  \, 1 \, {\rm GeV}) \simeq \frac{2}{3} [M^2(\Omega_Q^{(3/2)}) - M^2(\Omega_Q)  ] \sim 0.2$. 

\item 
However, HQE can be applied to charm baryons only qualitatively. Thus contributions both of ${\cal O}(1/m_c^2)$ and ${\cal O}(1/m_c^3)$  can have large impact, 
which is not surprising. Furthermore, other resonances could have large impact here, for which we have little understanding.

\item 
Only about $\tau(\Omega_c^0)$, but not on $\tau(\Lambda_c^+)$, $\tau(\Xi_c^+)$ \& $\tau(\Xi_c^0)$, while these four charm baryons carry spin 1/2? 
Well, the situations are different for $\Omega_c^0$ on one side, while $\Lambda_c^+$ and $\Xi_c^+$ \& $\Xi_c^0$ on the other side, namely: 
$\Omega_c^0 = [c(ss)_{j=1}]$ vs. $\Lambda_c^+ = [c(ud)_{j=0}]$ \& $\Xi_c = [c(sq)_{j=0}]$ with $q=u,d,s$. 
Again, resonances might have more impact, in particular for $\Omega_c^0$. 

\item
We have a `state-manager' for this `stage, namely to apply `duality' between hadrons and quarks that can be subtle. 
More comments about charm baryon wave functions had been given in Ref.\cite{BBD}.  
This item is not local in general: it means to compare the transitions of hadrons vs. quarks over 
a energy region $\sim 1 - 1.5$ GeV. The situations are different, when one looks at thresholds and resonances including broad ones. 

\item
Another test of non-local ``duality"? 

\end{itemize}
Lattice QCD will test our understanding of fundamental dynamics. However, we need more data about the 
lifetimes of charm baryons -- but also to measure the semi-leptonic widths of $\Xi_c^{+}$ / $\Xi_c^{0}$ \& $\Omega_c^0$ and discuss the results.

\subsection{Semi-leptonic widths of $\Lambda_c^+$, $\Xi_c^{+}$, $\Xi_c^{0}$ \& $\Omega_c^0$}
\label{SEMILEPTON}

The semi-leptonic width has been measured for $\Lambda_c^+$: 
\bea
{\rm BR}(\Lambda_c^+ \to e^+ \nu X)|_{\rm PDG2020} & =& (3.95 \pm 0.35)  \%  
\\
{\rm BR}(\Lambda_c^+ \to e^+ \nu X)/{\rm BR}(D^0 \to e^+ \nu X)|_{\rm PDG2020}& \simeq & 0.61 \pm 0.05 \; .  
\eea 
These values are consistent with HQE expectations.  They have not been measured yet for 
$\Xi_c^+$, $\Xi_c^0$ \& $\Omega_c^0$. It is very important to test our understanding of those. 
When one looks at the literature, one can see large different values. It had been estimated \cite{VOLO96}: 
\bea
\left.  \frac{{\rm BR}_{\rm SL}(\Xi_c^0)}{ {\rm BR}_{\rm SL}(\Lambda^+_c)} \right|_{\rm HQE} \sim 1  \; \; &\leftrightarrow &\; \; 
\left. \frac{ \tau(\Xi_c^0) }{\tau (\Lambda_c^+)} \right|_{\rm HQE} \sim 0.5
\\
\left. \frac{{\rm BR}_{\rm SL}(\Xi_c^+)}{ {\rm BR}_{\rm SL}(\Lambda^+_c)} \right|_{\rm HQE} \sim  2.5 \; \; &\leftrightarrow &\; \; 
\left. \frac{ \tau(\Xi_c^+) }{\tau (\Lambda_c^+)} \right|_{\rm HQE} \sim 1.7  
\\
{\rm BR}_{\rm SL}(\Omega_c^0)|_{\rm HQE} \; \; &\sim& \; \; (10 - 15) \; \% \; . 
\eea
As said above, one can compare the semi-leptonic widths of $\Omega_c^0 = [c(ss)_{j=1}]$ vs. $\Lambda_c^+ = [c(ud)_{j=0}]$.  
These values were based on a value of $\tau (\Omega_c^0)$ that is wrong as discussed in {\bf Sect.\ref{BARYONSLIFE}}.  
The correct value of ${\rm BR}_{\rm SL}(\Omega_c^0)$ should go {\em up} 
\footnote{One can find different values of ${\rm BR}_{\rm SL}(\Omega_c^0)$ based on baryon wave function $|\psi^{\Lambda_c^+} (0)|^2$ \cite{MELIC}; it had suggested 
this value go up to 25 \%; it has assumed $\mu^2_{\pi} \simeq 0.1 \; {\rm GeV}^2$, 
while $\mu^2_G (\Omega_c^0) \simeq 0.182 ({\rm GeV}^2$. However, we have learnt about QCD: $\mu^2_{\pi} \geq \mu^2_G$. There might different `roads' for such values.}. 
On the other hand, when one can follow the arguments in Ref.\cite{HYC2018} (like $\Gamma_{\rm SL} (\Omega_c^0) \sim \Gamma_{\rm SL} (\Lambda_c^+)$), one gets 
${\rm BR}_{\rm SL}(\Omega_c^0)|_{\rm HQE} \sim 6\; \% $. It is another test of non-local "duality".

The record of applying HQE for weak dynamics including charm baryons is good, when one is realistic:  
\bea 
\left. \frac{ \tau(\Xi_c^0) }{\tau (\Lambda_c^+)} \right|_{\rm HQE} \sim 0.5 \; \; \; \; &{\rm vs.}& \; \; \; \; 
\left.  \frac{ \tau(\Xi_c^0) }{\tau (\Lambda_c^+)} \right|_{\rm PDG2020} \simeq 0.76
\\
\left. \frac{ \tau(\Xi_c^+) }{\tau (\Lambda_c^+)}\right|_{\rm HQE} \sim 1.7  \; \; \; \; &{\rm vs.}& \; \; \; \;  
\left. \frac{ \tau(\Xi_c^+) }{\tau (\Lambda_c^+)} \right|_{\rm PDG2020} \simeq 2.25
\eea
There are two points: (a) Can our community measure those with data from LHCb experiment 
from run-2 (or run-3) -- or has to wait for data from Belle II? (b) On the theoretical side: which  branching ratios give us the best understanding about the underlying dynamics? 
As we had said before: there is no `golden medal'; one has to discuss the connections of these semi-leptonic branching ratios with their lifetimes!

Before we have cleared up our understanding of $C =1$ dynamics, we do not discuss here the situation about $C = 2$ baryons. 
However, we give a very short comment: the LHCb experiment has established the existence of the baryon $\Lambda^{++}_{\rm ccu}$ with a weak decay, 
although it had mostly focused on spectroscopy in $\Delta C =2$ dynamics.

\section{Studies of {\bf CP} asymmetries in charm transitions} 
\label{FUTURE}

Above we had discussed {\em inclusive} non-leptonic and semi-leptonic decays of charm hadrons. 
Again, we assume {\bf CPT} invariance; thus one has to go for {\bf CP} violation after {\em exclusive} decays, 
in particular for non-leptonic ones 
\footnote{PDG2020 has listed in a favored Cabibbo transition:   
$A_{\bf CP}(D^+ \to K_S\pi^+) = - (0.41 \pm 0.09)\%$, while $A_{\bf CP}(D^+ \to K_S\pi^+)|_{\rm SM} = -\, 2\, {\rm Re}\, \epsilon_K \simeq -\, 0.33\, \%$ 
due to {\em indirect} {\bf CP} in $K^0 - \bar K^0$ oscillation. Impact of ND could `hide' there, but we will not bet on that.}. 
{\bf CP} asymmetry in the decays of charm hadrons has been established for the first time: it is direct {\bf CP} violation in 
$D^0 \to K^+K^-$/$\pi^+\pi^-$ \cite{LHCbGuy}. It is the first step for a long `travel' about {\bf CP} asymmetries in charm transitions. 
The next steps are to find it in the decays of other charm hadrons, namely $D^+$ \& $D^+_s$ and $\Lambda_c^+$, $\Xi^{+}_c$, $\Xi^{0}_c$ \& $\Omega^0_c$; 
these six charm hadrons can produce only direct {\bf CP} violation. One has to remember there are two classes of non-leptonic transitions of charm hadrons 
to probe {\bf CP} violation, namely in singly Cabibbo suppressed (SCS) transitions (see $D^0 \to K^+K^-$/$\pi^+\pi^-$) and doubly Cabibbo suppressed (DCS) ones 
(like $D^0 \to K^+\pi^-$). In the latter case the SM can hardly produce there, see above and below. 
In Ref.\cite{CICERONE} from 2003 we had discussed {\bf CP} violation mostly about 2-body FS, but we had comments about 3- \& 4-body FS like Dalitz 
plots and {\bf T}-odd moments. Yet in 2019 the landscape has changed as said above, see Eq.(\ref{CharmCPV}): direct {\bf CP} asymmetry has been established 
in $D^0 \to K^+K^-/\pi^+\pi^-$ transitions, namely in SCS ones. In Ref.\cite{CICERONE} we had focused on indirect {\bf CP} violation. 
Now one has to change the strategy: direct {\bf CP} asymmetry has been established in one transition, namely $D^0 \to K^+K^-/\pi^+\pi^-$, 
see Eqs.(\ref{ADIR},\ref{AGAMMA},\ref{FIRST}). Unlikely to find non-zero value for $A_{\Gamma} (K^+K^-)$ `soon'.   
Next one can probe {\bf CP} asymmetry in $D^+_s \to K_S\pi^+/K^+\pi^0$ and $\Lambda_c^+ \to p\pi^0/\Lambda^0K^+$. There is a long list of 
2-body FS in $D$ decays, where one probe {\bf CP} asymmetries \cite{HYCCWC}. 
It is not clear whether the LHCb collaboration can join this competition about 2-body FS.

However, the FS of charm hadrons are mostly given by 3- \& 4-body FS, while 2-body ones are a small part 
\footnote{For favored Cabibbo, SCS \& DCS branching ratios the data give for 2-body FS with few $10^{-2}$, $10^{-3}$ \& $\sim 10^{-4}$, 
while with 3- \& 4-body ones as $\sim 0.25$, $\sim 0.01$ \& several$\times 10^{-4}$. As said above, one has more candidates for 
many-body FS.}. 
It is crucial to understand the information given by the data, namely to measure {\bf CP} asymmetries in 3- \& 4-body FS.  
To probe many-body FS in charm hadrons it gives experimental challenges in general; it is larger, since one has 
to deal with other backgrounds. In $pp$ collisions (like the LHCb experiment) one has to worry about production asymmetries, in particular for baryons. 
On the good side: in 3-body FS one can use well-known tools to deal with that, namely Dalitz plots; they are independent of production asymmetries.  
Actually, just looking at the data is not enough; one has to apply more refined like dispersion relations: they are above models, but below true QFT; 
their limits are $\sim$ 0.5 - 1.5 GeV. It needs some judgment, which resonances can contribute, including broad ones. 
The good site: one can apply to charm transitions, which is much better than for beauty transition. The bad site is, when 
one goes for small values. It is crucial to connect $D \to h_1\bar h_2 h_2$ with $D \to h_1\bar h_3 h_3$.
For 4-body FS one has to learn about more information with regional ones. 
These charm mesons have less observables than charm baryons. 
Often it is seen as a good sign, but not always.

\subsection{{\bf CP} asymmetries in SCS with many-body FS}
\label{CPVSCS}

Their amplitudes are described by effective operators $c\to u... \bar d ... d$ \& $c\to u... \bar s ... s$. 
The team of the LHCb collaboration, who has found direct {\bf CP} asymmetry, will continue to analyze run-2 data, 
namely to establish {\em indirect} {\bf CP} violation in $D^0 \to K^+K^-$ including {\em time depending} one, where one gets more information about the 
underlying dynamics; we will see. 

Our community has to  continue probing {\em direct} {\bf CP} asymmetries in 3- \& 4-body FS in the weak decays of charm mesons -- 
$D^+_{(s)}$ \& $D^0$ -- and charm baryons -- $\Lambda_c^+$, $\Xi_c^+$, $\Xi_c^0$ \& $\Omega_c^0$. 
For practical reasons we focus on 3-body FS in the decays of $D^+_{(s)}$ \& $\Lambda_c^+$ and 4-body FS of $D^0$ decays. 

\subsubsection{Dalitz plots for $D^+_{(s)}$}
\label{DALITZ1}

We list the decays that have been measured so far, see PDG2020:
\begin{itemize}
\item
BR$(D^+ \to \pi^+\pi^-\pi^+) = (3.27 \pm 0.18)\cdot 10^{-3}$ with ${\rm A_{\bf CP}} = (-\, 2 \pm 4)\cdot 10^{-2}$;

BR$(D^+ \to \pi^+K^-K^+) = (9.68 \pm 0.18)\cdot 10^{-3}$ with ${\rm A_{\bf CP}} = (+\, 0.37 \pm 0.29)\cdot 10^{-2} $.

\item
BR$(D^+_s \to K^+\pi^-\pi^+) = (6.5 \pm 0.4)\cdot 10^{-3}$ with ${\rm A_{\bf CP}} = (+ \,4 \pm 5)\cdot 10^{-2} $;

BR$(D^+_s \to K^+K^-K^+) = (0.218 \pm 0.020)\cdot 10^{-3}$, while no limit for  ${\rm A_{\bf CP}}$ is given.

\end{itemize}
Above we have talked about re-scattering in general in {\bf Sect.\ref{RESC}}, see Eq.(\ref{MANYBODY}) there.  Now we talk about the connections of Dalitz plots 
due to re-scattering, see just above. 

Dalitz plots could exhibit sizable asymmetries in different regions of 
varying signs that can largely cancel each other when one integrates over the whole phase space 
\footnote{We have an example from $B^{\pm} \to \pi^{\pm} \pi^+\pi^-$ \&  
$B^{\pm} \to \pi^{\pm} K^+ K^-$ or $B^{\pm} \to K^{\pm} \pi^+\pi^-$ \&  
$B^{\pm} \to K^{\pm} K^+ K^-$  of LHCb data from run-1, although at larger ratios.}. 
One expects {\em averaged} {\bf CP} asymmetries from the SM of ${\cal O}(10^{-3})$, while one can find {\em regional} ones of ${\cal O}(10^{-2})$ (or more). 

These transitions are analyzed by the LHCb collaboration from run-2 (and later will be from run-3) with smaller uncertainties. 
Fitting the data does not get the best information 
about the underlying dynamics; often one gets better ones based on {\em correlations} with other transitions: 
(a) Re-scattering $\pi \pi \leftrightarrow \bar K K$ both to $D^+$ \& $D^+_s$ due to non-perturbative QCD and (b)  connection between 
$D^+$ \& $D^+_s$. 

That can be probed using 
dispersion relations and chiral symmetry based on low energy data. Of course, data are the referees in the end; however, 
it needs some judgment which tools give us the best information with finite data. 
Simulations of these transitions had been discussed with details about 
impact of resonances $\rho$, $K^*$ \& $\phi$, but also  broad ones like $f_0(500)/\sigma$, $K_0^*(700)/\kappa$ etc. \cite{BEDIAGA} .

Real data will give more information about the underlying dynamics. 
Our community is waiting for LHCb results from run-2. Likewise for $D^+_s \to K^+K^-K^+$/$K^+\pi^-\pi^+$ and their connections from 
$D^+ \to \pi^+\pi^-\pi^+$/$\pi^+K^-K^+$, see Eq.(\ref{MANYBODY}). 

Of course, there are challenges both on the experimental and theoretical sides. 
In the latter case one has to think about the impact of non-perturbative QCD on our understanding about the measured {\bf CP} asymmetries and to compare 
the results in the decays of charm mesons vs. charm baryons.

\subsubsection{Dalitz plots for $\Lambda_c^+$ }
\label{DALITZ2}

Again, we compare the SCS transitions with favored one in 3-body FS:    
${\rm BR}(\Lambda_c^+ \to p\pi^+\pi^-) \simeq  4.6 \cdot 10^{-3}$  and ${\rm BR}(\Lambda_c^+ \to p K^+ K^-) \simeq   1.1 \pm \cdot 10^{-3}$ 
vs. ${\rm BR}(\Lambda_c^+ \to pK^-\pi^+) \simeq  6.3 \cdot 10^{-2}$ 
\footnote{It shows the impact of non-perturbative QCD due to differences from naive expectations of $\lambda^2 \sim 0.05$.}.  
From PDG2019 one gets  averaged direct {\bf CP} asymmetry: $\Delta A_{\bf CP} \equiv A_{\bf CP} (\Lambda_c^+ \to p K^+ K^-) -  A_{\bf CP}(\Lambda_c^+ \to p\pi^+\pi^-)  = (0.3 \pm 1.1) \cdot 10^{-2}$. 
It is unlikely that even ND could produce such values. On the other hand, the SM could produce {\em regional} {\bf CP} asymmetries of ${\cal O}(10^{-2})$, namely 
the impact of non-perturbative QCD due to resonances (including broad ones) like for $D^+$ ones: in one region one might find a {\bf CP} asymmetry with ${\cal O}(10^{-2})$, while 
in another region of ${\cal O}(10^{-2})$ with the opposite sign; to sum them one gets a {\bf CP} asymmetry of ${\cal O}(10^{-3})$. One can probe them,  if one has enough data to establish {\bf CP} asymmetry. It is another challenge if it shows the impact of ND.

\subsubsection{${\bf T}$-odd correlations in $D^0 \to K^+K^-\pi^+\pi^-$  (\& $D^0 \to \pi^+\pi^-\pi^+\pi^-$)} 
\label{DKKPIPI}

The next step is to discuss 4-body FS about averaged and regional {\bf CP} asymmetries. The landscapes of these FS are more complex due to the impact of 
non-perturbative QCD than with sums of  exclusive FS. We give two examples with their branching ratios from PDG2020: 
${\rm BR}(D^0 \to K^+K^-\pi^+\pi^-) \simeq 2.5 \cdot 10^{-3}$ (\& ${\rm BR}(D^0 \to \pi^+\pi^-\pi^+\pi^-) \simeq 7.6 \cdot 10^{-3}$).  

Assuming {\bf CPT} invariance {\bf T} and {\bf CP} asymmetries are the same meaning about underlying dynamics.
In the rest frames of $D^0$ \& $\bar D^0$ one can define $C_T \equiv \vec p_{K^+} \cdot (\vec p_{\pi^+} \times \vec p_{\pi^-})$ \&  
$\bar C_T \equiv \vec p_{K^-} \cdot (\vec p_{\pi^-} \times \vec p_{\pi^+})$. Under time reversal 
one gets both $C_T \to - C_T$ and $\bar C_T \to -  \bar C_T$; however, $C_T \neq 0$ does not necessarily establish {\bf T} violation; 
actually one get mostly non-zero values due to strong dynamics. 
{\bf CP} invariance tells us: $C_T = \bar C_T$; thus 
\beq
A_{\bf T} = \frac{1}{2} (C_T - \bar C_T ) \neq 0 \; . 
\label{CPVTR}
\eeq
The first step is to probe {\em T}-odd {\em momenta} for $D^0 \to K^+K^-\pi^+\pi^-$: 
\bea
\nonumber
\langle  C_T   \rangle = \frac{\Gamma_{D^0}(C_T > 0) - \Gamma_{D^0}(C_T < 0)}{\Gamma_{D^0}(C_T > 0) +  \Gamma_{D^0}(C_T < 0)}   &,& 
\langle  \bar C_T   \rangle = \frac{\Gamma_{\bar D^0}(\bar C_T < 0) - \Gamma_{\bar D^0}(\bar C_T > 0)}{\Gamma_{\bar D^0}(\bar C_T < 0) +  \Gamma_{\bar D^0}(\bar C_T > 0)} 
\\
\langle A_{\bf T} \rangle &=& \frac{1}{2} (\langle C_T\rangle - \langle  \bar C_T \rangle )
\eea 
In 2002 it was pointed out: $ \langle A_{\bf T}(D^0 \to K^+K^-\pi^+\pi^-)\rangle = (75 \pm 64 ) \cdot 10^{-3}$ \cite{CICERONE}; now PDG2020 gives: 
$\langle A_{\bf T}(D^0 \to K^+K^-\pi^+\pi^-)\rangle = +\, (2.9 \pm 2.2 ) \cdot 10^{-3}$. In the SM one expects a value for $A_{\bf T}$ of  ${\cal O}(10^{-3})$ 
for this SCS transition. 
\begin{enumerate}
\item
With more data the LHCb and Belle II collaborations have to probe semi-regional {\bf CP} asymmetries to go beyond as discussed in Ref.\cite{CICERONE}; a simple example: 
\bea
\frac{d\Gamma}{d\phi} (D^0 \to K^+K^-\pi^+\pi^-)&=& \Gamma_1\; {\rm cos}^2 \phi + \Gamma_2\; {\rm sin}^2 \phi +\Gamma_3 {\rm cos} \phi \;{\rm sin} \phi
\\
\frac{d\Gamma}{d\phi} (\bar D^0 \to K^+K^-\pi^+\pi^-)&=& \bar \Gamma_1\; {\rm cos}^2 \phi + \bar \Gamma_2\; {\rm sin}^2 \phi + \bar \Gamma_3 {\rm cos} \phi \;{\rm sin} \phi \; , 
\eea
where $\phi$ is the angle between the $K^+K^-$ and $\pi^+\pi^-$ planes. One gets six observables: {\bf CP} \& {\bf T} invariance leads to 
$\Gamma_1 = \bar \Gamma_1 $, $ \Gamma_2 = \bar \Gamma_2$ and $\Gamma_3 = - \bar \Gamma_3$.  
It is quite possible (or even likely) that a difference in $\Gamma_3$ vs. $\bar \Gamma_3$ is significantly larger than in $\Gamma_1$ vs. $\bar \Gamma_1$ and/or 
$\Gamma_2$ vs. $\bar \Gamma_2$. Furthermore one can expect that a differences in detection efficiencies can be handled by 
comparing $\Gamma_3$ with $\Gamma_{1,2}$ and $\bar \Gamma_3$ with $\bar \Gamma_{1,2}$. 
$\Gamma_3$ \& $\bar \Gamma_3$ constitute {\em T}-odd correlations. The moments of integrated forward-backward asymmetry lead to:
\beq
\langle A_{FB} \rangle |_0^{\pi} \simeq \frac{\Gamma_3 + \bar \Gamma_3 }{\pi(\Gamma_1 + \Gamma_2 + \bar \Gamma_1 + \bar \Gamma_2)}  \; . 
\label{AFB}
\eeq
PDG2019 has shown the analyses of $A_{\bf T}(D^0 \to [K^+K^-][\pi^+\pi^-])$ with the present data. As we had said before in Ref.\cite{CICERONE} we agree. 
However, the present data can give us more information about the underlying dynamics. There are two other different ways to define of two planes and the angle $\phi$ between them: 
\item
Or the angle $\phi$ is between the two planes $K^+\pi^-$ \& $K^-\pi^+$. There are several resonances that have impact; some are somewhat narrow  (like $K^*$), while they are broad (like $K^*_0(700)/\kappa$).

\item
Or the angle $\phi$ is between the two planes $K^+\pi^+$ \& $K^-\pi^-$. The phase shifts have not should resonant behavior so far. 

\end{enumerate}
Of course, it needs much more analyses to get this information to deal with experimental uncertainties. It is not trivial, but very important to understand the underlying dynamics. 

With much more data and more refined analyses one can to go {\em beyond}  $\langle A_{\bf T} \rangle$ and  $\langle A_{FB}\rangle$.  
Two examples: 
\begin{itemize}
\item
In general one can also compare $\Gamma_1$ vs. $\bar \Gamma_1$ and $\Gamma_2$ vs. $\bar \Gamma_2$. 

\item
The second example is much complex, when one probes {\bf CP} asymmetries in $D^0 \to \pi^+\pi^-\pi^+\pi^-$. How can 
somebody differentiate  between the two $\pi^+$ and $\pi^-$? In the neutral $D$ rest frame one can call one is fast, while the other 
is slow. However, it does not mean the quantitive patterns in $\pi^+_{\rm fast}$ \& $\pi^+_{\rm slow}$ vs. $\pi^-_{\rm fast}$ \& 
$\pi^-_{\rm slow}$  are the same. The best fitted results often do not give the best understanding of the underlying dynamics -- 
it needs `judgement' based in connection with other transitions. 

\end{itemize}
One can see a more subtle example in Appendix in {\bf Sect.\ref{SEGHAL}}.

\subsection{{\bf CP} asymmetries in DCS} 
\label{CPVDCS}

Their amplitudes are described by an effective operator $c\to u... \bar s ... d$. 
The SM gives basically zero-values of {\bf CP} asymmetries for DCS as discussed above. On the good side: they are 
`hunting' regions for ND. However, there are also bad sides: (a) `We' have to wait for future data from the LHCb experiment of runs-3/4 or data from Belle II. 
(b) The rates of DCS are very small; there is a true challenge to deal with the backgrounds in the data.  Again, 3- \& 4-body FS are larger than 2-body ones, 
see below. Furthermore, the  branching ratios of DCS decays of $D^0$, $D^+_{(s)}$ and $\Lambda_c^+$ are much smaller than naive scale  
$\lambda^2 = (0.223)^2 \simeq 0.05$; it shows the impact of non-perturbative QCD.

\subsubsection{Charm mesons}
\label{CPVDCSMESON}

The landscape of 2-body FS for clear DCS transitions are thin: $D^0 \to K^+\pi^-$ and $D^+ \to K^+ \pi^0$/$\eta$/$\eta^{\prime}$, while  
nothing  for $D^+_s$. One can compare them with CF ones: 
\bea
{\rm BR}(D^0 \to K^+\pi^-) \sim 1.4 \cdot 10^{-4} \; \; &{\rm vs.}& \;  \;  {\rm BR}(D^0 \to K^-\pi^+) \sim 4 \cdot 10^{-2}
\\ 
{\rm BR} (D^+ \to K^+\pi^0/\eta/\eta^{\prime}) \sim 5 \cdot 10^{-4} \; \; & {\rm vs.} &  \; \;  {\rm BR} (D^+ \to K_S \pi^+) \sim 1.6 \cdot 10^{-2} \; . 
\eea
DCS branching ratios are sizably smaller than favored Cabibbo ones based on naive scale $\lambda^2 = (0.223)^2 \simeq 0.05$. 

DCS transitions are affected by oscillations as said  before: 
\bea
\nonumber
\Gamma (D^0(t) \to K^+\pi^-) \propto |(A(D^0 \to K^+\pi^-)|^2 \left[ 1+
\left( \frac{t}{\tau_D} \right) ^2 \left( \frac{x_D^2 + y_D^2}{4}
\left| \frac{q}{p}\bar \rho (K^+\pi^-)   \right|^2   \right) -     \right.
\\
\left.  -\left(\frac{t}{\tau_D} \right) \left[ y_D {\rm Re}\left( \frac{q}{p}\bar \rho (K^+\pi^-) \right)
+x_D {\rm Im}\left( \frac{q}{p}\bar \rho (K^+\pi^-)  \right)  \right]  \right]
\eea
\bea
\Gamma (\bar D^0 (t) \to K^-\pi^+) \propto |A(\bar D^0 \to K^-\pi^+)|^2 \left[ 1+  \left( \frac{t}{\tau_D} \right)^2  \left(\frac{x_D^2 + 
y_D^2}{4}\left| \frac{p}{q}\rho (K^-\pi^+)   \right|^2 \right) -  \right.
\\
\left.  -  \left( \frac{t}{\tau_D} \right)  \left[ y_D {\rm Re}\left( \frac{p}{q} \rho (K^-\pi^+) \right) 
+ x_D {\rm Im}\left( \frac{p}{q}\rho (K^-\pi^+) \right) \right]  \right]
\eea
We know that that $D^0 - \bar D^0$ oscillations are slow -- $x_D$, $y_D$ $\ll$ 1;  
\beq
A_{\bf CP}(D^0(t) \to K^+\pi^-) \simeq A_{\bf CP}(D^0 \to K^+\pi^-) - \frac{t}{\tau (D^0)} |\bar \rho (K^+\pi^-)| {\rm sin}\phi_{K\pi}  \
; , 
\label{CPVD0K+PI-}
\eeq
where we use the following notation:
\bea
\frac{q_D}{p_D} \bar \rho (K^+\pi^-) &\equiv& - |\bar \rho (K^+\pi^-)|e^{-i(\delta - \phi_{K\pi})} 
\\
\frac{p_D}{q_D} \rho (K^-\pi^+) &\equiv& - | \rho (K^-\pi^+|e^{-i(\delta + \phi_{K\pi})} \; , 
\eea
with $\delta$ and $\phi_{K\pi}$ the strong and weak phases, respectively.  
As said above, in DCS the SM gives basically zero value, while ND could produce non-zero values. 
Obviously time-dependent analyses are subtle, but very important.
Present data about {\bf CP} asymmetry has led to in time-independent analysis:
\beq
A_{\bf CP} (D^0 \to K^+\pi^-) = (-\, 0.9 \pm 1.4 ) \, \% \; ; 
\eeq
i.e., it is not close to values, what one can hope.  

However, we make a general point as said before and above: our community has to probe {\bf CP} violation in many-body FS -- in particular in 
3- \& 4-body FS -- both on global and regional ones. Look at the present data of branching ratios by comparing DCS vs. favored ones: 
\begin{itemize}
\item 
BR$(D^0 \to K^+\pi^-\pi^0)\sim 3.1 \cdot 10^{-4}$ vs. BR$(D^0 \to K^-\pi^+\pi^0) \sim 14.4 \cdot 10^{-2}$

BR$(D^0 \to K^+\pi^+\pi^-\pi^-) \sim 2.5 \cdot 10^{-4}$ vs. BR$(D^0 \to K^-\pi^+\pi^+\pi^-)\sim 8.2 \cdot 10^{-2}$

\item
BR$(D^+ \to K^+\pi^+\pi^-) \sim 4.9 \cdot 10^{-4}$ vs. BR$(D^+ \to K^-\pi^+\pi^+) \simeq  9.4  \cdot 10^{-2} $

\item
BR$(D^+_s \to K^+K^+\pi^-) \sim 2.1 \cdot 10^{-4}$ vs. BR$(D^+_s \to  K^+K^-\pi^+) \sim 5.4 \cdot 10^{-2} $

\end{itemize}
PDG2020 has not given even limits for BR$(D^0 \to K^+K^-K_S)$  or BR$(D^0 \to K^+K^+K^-\pi^-)$.

\subsubsection{Charm baryons}
\label{CPVDCS}

So far, the present data of DCS decays calibrated by favored one are very thin, namely only one:  
\beq
{\rm BR}(\Lambda_c^+ \to p K^+\pi^-) \sim 1.11 \cdot 10^{-4} \; \; {\rm vs.} \; \; {\rm BR}(\Lambda_c^+ \to p K^-\pi^+) \sim 6.3 \cdot 10^{-2}
\eeq
If one has establish {\bf CP} asymmetry in $\Lambda_c^+ \to p K^+\pi^-$, one has found the existence of ND in charm decays.

Therefore we will not discuss the decays of $\Xi_c^+$, $\Xi_c^0$ or even $\Omega_c^0$ here. 
However for the future with the run-3 (\& run-4), it would give us now very important lessons about underlying dynamics.

\section{Summary} 
\label{SUM}

The end of run-2 of the LHCb experiment had happened now; run-3 will start in 2021 for around three years and the next era will hopefully start at 2025. 
`Soon' the Belle II collaboration will enter to get new information about heavy flavor dynamics.

In 2020 we have gotten new information about two classes of charm transitions: 
\begin{itemize} 
\item
The lifetimes and semi-leptonic branching ratios of charm hadrons show the impact of non-perturbative forces. 
They are connected, but they are not straightway. Measured semi-leptonic branching ratios give us tests of the ability  
of an experimental collaboration to produce data, but also to understand the underlying dynamics. It is crucial to analyze  
different transitions; i.e., one should not focus on a `golden' test. One has to think more about applying OPE and HQE to QCD. 

\item
{\bf CP} asymmetries in charm  transitions depend on the connections of QCD and weak dynamics -- including possible indirect impact of ND. It is a great 
achievement of the LHCb collaboration, but we are just at the beginning of a long travel.

\end{itemize}
Obviously the `actors' in this `dramas' are not the same; however, they are connected in subtle ways.

\subsection{Inclusive transitions of charm hadrons}
\label{SUMLIFES}

HQE based on OPE has been very successful in describing {\em quantitatively} the weak lifetimes and semi-leptonic transitions of beauty hadrons, 
see Eq.(\ref{HEAVYWIDTH}). 

\subsubsection{Understanding dynamics of lifetimes for charm hadrons}
\label{LIFETIMESFUTURE}

It is amazing how HQE can be applied to the charm mesons $D^+_{(s)}$ \& $D^0$ already semi-quantitatively with a factor even 2.5. Can one apply it to charm baryons 
at least qualitatively? Through 2018 data were seen as `natural' also for charm baryons, in particular for $\Omega_c^0$ as discussed above. 
However, the `landscape' of charm baryons has sizably been changed in 2019/20 with $\tau(\Omega_c^0)$ and $\tau(\Xi_c^0)$ (\& more). 
\bea
\tau (\Xi_c^+)/ \tau (\Xi_c^0) |_{\rm PDG2020} \sim  2.98  \; &,& \;  
\tau (\Xi_c^+)   / \tau (\Xi_c^0) |_{\rm LHCb, run-1} \simeq 2.96
\\
\tau (\Xi_c^+)/ \tau (\Xi_c^0) |_{\rm HQE} &\sim&  2.8 
\\
\tau (\Xi_c^0)/\tau (\Lambda_c^+)  |_{\rm PDG2020} \simeq 0.76 
\;  &,& \;  
\tau (\Xi_c^0) / \tau (\Lambda_c^+)  |_{\rm LHCb, run-1} \simeq 0.76
\label{XIC0EXP}
\\
\tau (\Xi_c^0)/\tau (\Lambda_c^+)  |_{\rm HQE} &\sim &  0.5 
\label{XIC0THEO}
\\
\tau (\Xi_c^+)/\tau (\Lambda_c^+)  |_{\rm PDG2020} \simeq 2.25 
\; \; &,& 
\tau (\Xi_c^+)/\tau (\Lambda_c^+)  |_{\rm LHCb, run-1} \simeq 2.24 
\\
\tau (\Xi_c^+)/\tau (\Lambda_c^+)  |_{\rm HQE} &\sim & 2.2 \; . 
\label{SMALLTHEO} 
\eea 
There is some disagreement between Eq.(\ref{XIC0EXP}) vs. Eq.(\ref{XIC0THEO}). It could be solved by LHCb data 
from run-2 like $\tau (\Xi_c^0) / \tau (\Lambda_c^+)  |_{\rm LHCb, run-2} \simeq 0.66$. 
There are two classes of $\bf WS$ diagrams for $\Xi_c^0$: (a) Cabibbo favored one $[cds] \to (s...u...s)$, while 
(b) SCS ones $[csd]\to (s...u...d)$ \& $[cds] \to (d...u...s)$. When one is `realistic', HQE can cover that situation. 
Above we have pointed out the obvious weak part of applying HQE 
for charm transitions, in particular for charm baryons: so far quark models are used, not QCD. There are two less obvious weak parts including "duality" 
as we had discussed above.


\subsubsection{Semi-leptonic decays of charm hadrons}
\label{SLBRFUTURE}

As we have said above the measured semi-leptonic branching ratios of $D^+_{(s)}$ \& $D^0$ are well described by HQE in a semi-quantitative way. 
PDG2020 lists only measured ${\rm BR}(\Lambda_c^+ \to e^+ \nu X) = (3.95 \pm 0.35)\%$, while not for $\Xi_c^+$, $\Xi_c^0$ \& $\Omega_c^0$ ones. 
When semi-leptonic branching ratios are measured for $\Xi_c^0$ \& $\Xi_c^+$ 
-- even better including for $\Omega_c^0$ -- one can discuss the connections between the lifetimes and semi-leptonic branching ratios: 
they would lead to semi-quantitative predictions.    
It will improve with more data, analyses \& thinking: it will get better understanding of non-perturbative QCD  including the item of "duality". 

\subsection{Beginning of probing {\bf CP} asymmetries in charm hadrons}
\label{BEGINCPV}

Direct {\bf CP} asymmetry $\Delta A_{\rm CP}(D^0 \to K^+K^-/\pi^+\pi^-)$ has been established, 
which shows real progress about weak decays of charm hadrons. We had said above several times: our community is still in the beginning of a 
long travel about {\bf CP} asymmetries. It is an excellent achievement by the LHCb collaboration, and it opened a novel door  -- but it is the beginning! 
The next step is to go after indirect {\bf CP} violation in $D^0 \to K^+K^-$, basically due to experimental reasons. Of course, the LHCb experiment also 
goes after {\bf CP} asymmetry in $D^0 \to \pi^+\pi^-$. Yet one is still close to the beginning of this traveling, where these are only a small 
parts of $D^0$ transitions. One has to go after many-body FS, namely 3- \& 4-body FS of $D^0$, $D^+$ \& $D_s^+$ and many-body FS for charm baryons, 
as discussed in {\bf Sect.{\ref{CPVSCS}}} \& {\bf Sect.{\ref{CPVDCS}}}.

\subsection{Goals for `understanding' charm dynamics}
\label{GOALS}

More data, more analyses and more thinking are needed about the connections for several transitions in different directions and different levels. 
In the second decade of this century we have entered a new era and will continue in the third decade from LHCb and Belle II. It means that the 
theory community has to use refined tools to understand the information given by the data and to focus on the connections between transitions. 
\begin{itemize}
\item
Lifetimes and semi-leptonic decays of $D^+$, $D^0$ \& $D_s^+$ and $\Lambda_c^+$, $\Xi_c^+$, $\Xi_c^0$ \& $\Omega_c^0$ give new lessons 
about non-perturbative QCD.  The predictions are semi-quantitative for the charm mesons and $\Lambda_c^+$, while qualitative at least for 
$\Xi_c^+$ \& $\Xi_c^0$. We pointed out why previous prediction are wrong and to test it. 

\item
It is crucial to probe {\bf CP} asymmetries in 3-  \& 4-body FS of $D^+$, $D^0$ \& $D_s^+$ and $\Lambda_c^+$, $\Xi_c^+$ \&  $\Xi_c^0$; 
it would be good to continue these studies to $\Omega_c^0$.  

\item 
Our community has to wait for the results of run-3 of LHCb and Belle II to establish {\bf CP} asymmetries in DCS ones. 

\end{itemize}
Non-perturbative QCD has large impact on both items: sometimes it is obvious, while others are subtle.  Charm transitions are part of the much  larger 
landscape of the known matter. It will be discussed in a book to be published soon \cite{BRP}.

\vspace*{2mm}

{\bf Acknowledgments:}  We have appreciated the comments \& discussions from our colleague Hai-Yang Cheng: we missed one important paper 
from him about $\tau (\Omega_c)$; furthermore it helped to make it clearly where we agree or disagree. We have gotten short comments from 
Michael Morello about continuing project about time-dependent {\bf CP} violation \cite{MICHAEL}. 

This work was supported by the Italian Institute for Nuclear Physics (INFN) and the NSF PHY-1820860.


\appendix 
\section{Prediction of {\em large} {\bf CP} asymmetry in $K_L \to \pi^+\pi^- e^+e^-$} 
\label{SEGHAL}

Seghal had pointed out that the very rare $K_L \to \pi^+\pi^- e^+e^-$ (with a branching ratio with $\simeq 3 \cdot 10^{-7}$) should show large 
{\bf CP} asymmetry, namely a value of  $\sim 0.14$ based on $\epsilon_K \simeq 0.002$ in the SM \cite{SEGHAL92}. Indeed 
PDG2019 gives: $A_{\bf CP}(K_L \to \pi^+\pi^- e^+e^-)= 0.137 \pm 0.015$. 

We are {\em not} talking about history; one can analyze it in different situations including non-leptonic decays.  
Unit vectors aid in discussing this scenario in details, where 
\bea
\nonumber 
\vec n_{\pi} \equiv \frac{ \vec p_+ \times \vec p_- }{| \vec p_+ \times \vec p_-| } \; &,& \; 
\vec n_{l} \equiv \frac{ \vec k_+ \times \vec k_- }{| \vec k_+ \times \vec k_-| } 
\\
\vec z &\equiv & \frac{\vec p_+ + \vec p_- }{| \vec p_+ + \vec p_-   |}
\\
\nonumber
{\rm sin}\, \phi &=&  (\vec n_{\pi} \times \vec n_l  ) \cdot \vec z \, [{\bf CP}= -, {\bf T} = - ]
\\
{\rm cos} \, \phi &=& \vec n_{\pi} \cdot  \vec n_l   \, [{\bf CP}= +, {\bf T} = + ]
\\
\frac{d\Gamma }{d\phi } & \sim & 1 - (Z_3 \; {\rm cos} \, 2\phi + Z_1 \; {\rm sin} \, 2\phi  )
\eea
One can measure asymmetry in the momenta:
\beq
A_{\phi} = \frac{[\int_0^{\pi/2} d\phi - \int_{\pi/2}^{\pi} d\phi]\frac{d\Gamma}{d\phi} }{[\int_0^{\pi/2} d\phi + \int_{\pi/2}^{\pi}d\phi  ] 
\frac{d\Gamma}{d\phi}}   \; .
\eeq
There is an obvious reason for probing the angle between the two $\pi^+\pi^-$ \& $e^+e^-$. However, the situations are more complex for four 
pseudo-scalars hadrons:
\bea
\frac{d}{d\phi}\Gamma (H_Q \to h_1h_2h_3h_4)&= & |c_Q |^2 - [b_Q (2 \, {\rm cos}^2 \, \phi - \, 1) + 2 \, a_Q \, {\rm sin} \, \phi \; {\rm cos} \, \phi ]
\\
\frac{d}{d\phi}\Gamma (\bar H_Q \to \bar h_1 \bar h_2 \bar h_3 \bar h_4) &= & |\bar c_Q |^2 - [\bar b_Q (2 \, {\rm cos}^2 \, \phi - \, 1) + 2 \, \bar a_Q \, {\rm sin} \, \phi \; {\rm cos} \, \phi ]
\\
\Gamma (H_Q \to h_1h_2h_3h_4)  &= & |c_Q |^2 
\\
\Gamma (\bar H_Q \to \bar h_1 \bar h_2  \bar h_3  \bar h_4) &= & |\bar c_Q |^2  
\\
\langle  A_{\bf CP}\rangle |_0^{\pi} & = & \frac{2 (a_Q - \bar a_Q  )  }{|c_Q |^2 + |\bar c_Q |^2  } \; ;
\label{AQ}
\eea
i.e., the  terms $b_Q$ \& $\bar b_Q$ have no impact here; see the details in Ref.\cite{TIM}. 
In the future one can compare the results of Eq.(\ref{AFB})   \& Eq.(\ref{AQ})

\end{document}